\documentclass{aastex62}

\received{}
\revised{}
\accepted{}
\submitjournal{ApJ}

\shorttitle{Survey Observations of HC$_{3}$N and N$_{2}$H$^{+}$}
\shortauthors{Taniguchi et al.}
\begin{document}

\title{Survey Observations to Study Chemical Evolution from High-Mass Starless Cores to High-Mass Protostellar Objects II. HC$_{3}$N and N$_{2}$H$^{+}$}

\correspondingauthor{Kotomi Taniguchi}
\email{kt8pm@virginia.edu}

\author[0000-0003-4402-6475]{Kotomi Taniguchi}
\altaffiliation{Virginia Initiative on Cosmic Origins Fellow}
\affiliation{Departments of Astronomy and Chemistry, University of Virginia, Charlottesville, VA 22904, USA}

\author[0000-0003-0769-8627]{Masao Saito}
\affiliation{National Astronomical Observatory of Japan (NAOJ), National Institutes of Natural Sciences (NINS), Osawa, Mitaka, Tokyo 181-8588, Japan}
\affiliation{Department of Astronomical Science, School of Physical Science, SOKENDAI (The Graduate University for Advanced Studies), Osawa, Mitaka, Tokyo 181-8588, Japan}

\author{T. K. Sridharan}
\affiliation{Harvard-Smithsonian Center for Astrophysics, 60 Garden Street, MS 78, Cambridge, MA 02138, USA}

\author[0000-0001-9778-6692]{Tetsuhiro Minamidani}
\affiliation{Nobeyama Radio Observatory, National Astronomical Observatory of Japan (NAOJ), National Institutes of Natural Sciences (NINS), Nobeyama, Minamimaki, Minamisaku, Nagano 384-1305, Japan}
\affiliation{Department of Astronomical Science, School of Physical Science, SOKENDAI (The Graduate University for Advanced Studies), Osawa, Mitaka, Tokyo 181-8588, Japan}



\begin{abstract}
We have carried out survey observations of molecular emission lines from HC$_{3}$N, N$_{2}$H$^{+}$, CCS, and {\it {cyclic}-}C$_{3}$H$_{2}$ in the 81$-$94 GHz band toward 17 high-mass starless cores (HMSCs) and 28 high-mass protostellar objects (HMPOs) with the Nobeyama 45-m radio telescope.
We have detected N$_{2}$H$^{+}$ in all of the target sources except one and HC$_{3}$N in 14 HMSCs and in 26 HMPOs.
We investigate the $N$(N$_{2}$H$^{+}$)/$N$(HC$_{3}$N) column density ratio as a chemical evolutionary indicator of massive cores.
Using the Kolmogorov-Smirnov (K-S) test and Welch's t test, we confirm that the $N$(N$_{2}$H$^{+}$)/$N$(HC$_{3}$N) ratio decreases from HMSCs to HMPOs.
This tendency in high-mass star-forming regions is opposite to that in low-mass star-forming regions.
%The HC$_{3}$N column density increases from HMSCs to HMPOs, which is consistent with our previous results.
Furthermore, we found that the detection rates of carbon-chain species (HC$_{3}$N, HC$_{5}$N, and CCS) in HMPOs are different from those in low-mass protostars.
The detection rates of cyanopolyynes (HC$_{3}$N and HC$_{5}$N) are higher and that of CCS is lower in high-mass protostars, compared to low-mass protostars.
We discuss a possible interpretation for these differences.
\end{abstract}



\section{Introduction} \label{sec:intro}

Around 200 molecules have been detected in the interstellar medium or circumstellar shells so far\footnote{https://www.astro.uni-koeln.de/cdms/molecules} and the number of detected molecules has increased.
Chemical composition contains various information in star/planet forming regions including the past conditions and changes along the star formation processes \citep[e.g.,][]{2012A&ARv..20...56C}. 
There have been many studies working on the chemical evolution in low-mass star-forming regions.

\citet{1992ApJ...392..551S} carried out survey observations of CCS, HC$_{3}$N, HC$_{5}$N, and NH$_{3}$ toward 49 cores in the Taurus and Ophiuchus regions with the Nobeyama 45-m telescope.
The column density ratio of $N$(CCS)/$N$(NH$_{3}$) was suggested as a possible indicator of cloud evolution and star formation in low-mass star-forming regions.
\citet{2009ApJ...699..585H} also carried out survey observations of CCS, HC$_{3}$N, and HC$_{5}$N toward 40 cores and analyzed combining with the previous results. 
They confirmed that the abundance ratios of carbon-chain molecules and NH$_{3}$, in particular the $N$(NH$_{3}$)/$N$(CCS) ratio, are good indicators of chemical evolutionary stage of dark cloud cores: the $N$(NH$_{3}$)/$N$(CCS) ratio tends to be high in star-forming cores but low in starless cores.
\citet{1998ApJ...506..743B} carried out survey observations of N$_{2}$H$^{+}$, {\it {cyclic}}-C$_{3}$H$_{2}$({\it {c}}-C$_{3}$H$_{2}$), and CCS toward 60 dense cores with the 37 m telescope of the Haystack Observatory.
They found that the $N$(N$_{2}$H$^{+}$)/$N$(CCS) ratio is lower in starless cores than in star-forming cores by a factor of 2.
On the other hand, there was no significant difference in the $N$({\it {c}}-C$_{3}$H$_{2}$)/$N$(N$_{2}$H$^{+}$) ratio between starless and star-forming cores.
Moreover, a chemical evolution factor (CEF) was newly proposed \citep{2017ApJS..228...12T}.
The CEF is a parameter to represent the chemical evolution by using molecular column density ratios, and expected to distinguish starless and star-forming cores.
In summary, the above studies suggest that the ratios of $N$(carbon-chain species)/$N$(nitrogen-baring species) decreases with evolution.

The decreases in the $N$(carbon-chain species)/$N$(nitrogen-baring species) ratios can be explained by chemical characteristics of carbon-chain molecules and nitrogen-bearing species.
Carbon-chain molecules such as CCS and HC$_{2n+1}$N ($n = 1,2,3,...$) are formed from ionic carbon (C$^{+}$) and atomic carbon (C) at the early stage of molecular clouds and thus generally abundant in the early stage of chemical evolution.
On the other hand, NH$_{3}$ and N$_{2}$H$^{+}$ are abundant in the late stage of molecular clouds and their distributions are centrally condensed \citep[e.g.,][]{1996ApJ...468..761K}, because they are formed from N$_{2}$, whose formation is slow in dark clouds \citep{2017iace.book.....Y}.

Recent studies showed evidence that the solar system was born in a cluster, which resembles high-mass star-forming regions \citep[e.g.,][]{2010ARA&A..48...47A}.
Hence, tracing the chemical evolution in high-mass star-forming regions will lead to our understanding of the formation process of complex organic molecules including amino acids detected in meteorites or comets in the solar system.
Despite its importance, chemistry in high-mass star-forming regions, in particular the early stage, is still unclear and thus challenging topics.

The largest sample of molecular emission lines in high-mass star-forming regions was provided by the Millimeter Astronomy Legacy Team 90 GHz Survey \citep[MALT90;][]{2011ApJS..197...25F, 2013PASA...30...57J}.
Using the MALT90 data, there are several studies working on the chemical evolution in high-mass star-forming regions \citep[e.g.,][]{2013ApJ...777..157H, 2015MNRAS.451.2507Y, 2016PASA...33...30R}.
However, no clear chemical evolutionary indicator has been established yet in high-mass star-forming regions.

\citet{2008ApJ...678.1049S} carried out survey observations of N$_{2}$H$^{+}$, HC$_{3}$N, CCS, NH$_{3}$, and CH$_{3}$OH toward 55 massive clumps associated with infrared dark clouds.
The N$_{2}$H$^{+}$ emission lines were detected in most of the sources, whereas the CCS lines were not detected in any source.
The $N$(CCS)/$N$(N$_{2}$H$^{+}$) ratios are lower than unity even in the {\it {Spitzer}} 24 $\mu$m dark objects. 
Based on the results, \citet{2008ApJ...678.1049S} suggested that most of the massive clumps are chemically more evolved than those of the low-mass starless cores.
Changes in the ratios between carbon-chain species and nitrogen-bearing molecules have not been revealed in high-mass star-forming regions yet, as there was no CCS detection.

In order to study the chemical evolution in the early stage of high-mass star-forming regions, we have conducted survey observations of molecular emission lines toward high-mass starless cores (HMSCs) and high-mass protostellar objects (HMPOs) with the Nobeyama 45-m radio telescope.
\citet{2002ApJ...566..931S} identified 69 HMPOs, using far-infrared, radio continuum, and molecular data. 
In HMPOs, millimeter dust continuum emission was detected from all of the sources, whereas weak or no continuum emission at 3.6 cm was detected.
\cite{2005ApJ...634L..57S} identified 56 HMSCs, comparing images of fields containing candidate HMPOs at 1.2 mm and mid-infrared (MIR; 8.3 $\mu$m). 
HMSC was defined as a core showing 1.2 mm emission and absorption or no emission at the MIR wavelength suggestive of cold dust.
HMSCs are very likely in an earlier stage than HMPOs and may be with lower mass stars.

In this paper, we report results of the survey observations of HC$_{3}$N ($J$ = 9$-$8 and 10$-$9), N$_{2}$H$^{+}$ ($J$ = 1$-$0), CCS ($J_{N}$ = $6_{7}-5_{6}$), and {\it {para}}-{\it {cyclic}}-C$_{3}$H$_{2}$ ($J_{Ka, Kc}$ = $2_{0,2} - 1_{1,1}$) toward 17 HMSCs and 28 HMPOs with the Nobeyama 45-m radio telescope.
We select the 81 -- 94 GHz band so that we can derive the excitation temperatures, optical depths, and column densities of HC$_{3}$N and N$_{2}$H$^{+}$. 
This is the second season of the high-mass survey project with this telescope \citep[first season;][]{2018ApJ...854..133T}.
The main purpose of this survey project is to investigate the chemical evolution of massive stars from the early stage, i.e., starless core phase.
In the previous paper, we reported the survey observations of HC$_{3}$N ($J$ = 5$-$4) and HC$_{5}$N ($J$ = 16$-$15) in the 42$-$45 GHz band toward 17 HMSCs and 35 HMPOs.
They detected HC$_{3}$N from 15 HMSCs and 28 HMPOs and HC$_{5}$N from 5 HMSCs and 14 HMPOs, respectively.
They suggested that HC$_{3}$N is newly formed at HMPO stage in the warm dense gas where CH$_{4}$ and C$_{2}$H$_{2}$ are evaporated from grain mantles, using the statistical analyses.

We describe observations in Section \ref{sec:obs} and summarize observational results in Section \ref{sec:res}.
Analyzing method and the results are summarized in Section \ref{sec:ana}.
We compare the line widths among species (Section \ref{sec:disline}) and detection rates of carbon-chain species between HMPOs and low-mass protostars (Section \ref{sec:detection_rate}).
In Section \ref{sec:evolution}, we investigate the $N$(N$_{2}$H$^{+}$)/$N$(HC$_{3}$N) ratio as a chemical evolutionary indicator of massive cores.

\section{Observations} \label{sec:obs}

We carried out observations of molecular emission lines of HC$_{3}$N, N$_{2}$H$^{+}$, CCS, and {\it {c}-}C$_{3}$H$_{2}$ in the 81$-$94 GHz band simultaneously with the Nobeyama 45-m radio telescope during 2016 December, 2017 February and March (Proposal ID: 4163004, PI: Kotomi Taniguchi, 2016 -- 2017 season).
Target line, rest frequency, and excitation energy are summarized in Table \ref{tab:tab1}.
Target sources (17HMSCs and 28 HMPOs) were the same ones as in our first season of observations \citep{2018ApJ...854..133T}, which were selected from the HMSC source list \citep{2005ApJ...634L..57S} and the HMPO source list \citep{2002ApJ...566..931S} with the following characteristics:
\begin{enumerate}
\item The source declination is above $-6\arcdeg$ for HMSCs and $+6\arcdeg$ for HMPOs.
\item NH$_{3}$ has been detected.
\item HMPOs located in the same regions as the observed HMSCs ($-6\arcdeg <$ decl. $< +6\arcdeg$).
\end{enumerate}
We excluded 7 HMPOs, where \citet{2018ApJ...854..133T} carried out observations but no HC$_{3}$N emission line was detected (18437-0216, 18454-0158, 19403+2258, 19471+2641, 20081+2720, 22551+6221, 23545+6508).

The on-source positions were the same as ones where \citet{2018ApJ...854..133T} observed.
The coordinate of the observed positions were summarized in \citet{2005ApJ...634L..57S, 2002ApJ...566..931S}.
The off-source positions were set at the same positions with \citet{2018ApJ...854..133T}; no IRAS 100 $\mu$m emission positions\footnote{We used the SkyView (https://skyview.gsfc.nasa.gov/current/cgi/query.pl).} or low extinction ($A_{\rm {V}} < 1$ mag) positions\footnote{We used the all-sky visual extinction map generated by \citet{2005PASJ...57S...1D} (http://darkclouds.u-gakugei.ac.jp).}.
The position-switching mode with the chopper-wheel calibration method was employed.
We set the scan pattern at 20 s each for on-source and off-source positions.

We used the TZ receiver in the 2SB mode.
The beam size and main beam efficiency ($\eta_{\rm {mb}}$) were $18\arcsec$ and 54\%, respectively.
We used the SAM45 FX-type digital correlator in frequency setup whose bandwidth and resolution are 500 MHz and 122.07 kHz, respectively.
The frequency resolution of 122.07 kHz corresponds to 0.4 km s$^{-1}$.
We conducted 2-channel binning in the final spectra, which means that the velocity resolution of the final spectra is 0.8 km s$^{-1}$.
The system temperatures were between 140 and 210 K, depending on weather conditions and the elevation.

The telescope pointing was checked every 1$-$3 hr by observing the SiO maser lines ($J=1-0$) from U-Aur, RR-Aql, R-Aql, UX-Cyg, IRC+60334, and R-Cas.
We used the H40 receiver for pointing observations. 
The pointing error was within $3\arcsec$.

%%%%%%%%%%%%%%%%%%%%%%%%%%%%%%%%%%%%%%%%%%%%%%%%%%%%%%%%%%%%%%%%%
\floattable
\begin{deluxetable}{llcr}
\tabletypesize{\scriptsize}
\tablecaption{Summary of target lines \label{tab:tab1}}
\tablewidth{0pt}
\tablehead{
\colhead{Species} &  \colhead{Transition} & \colhead{Frequency\tablenotemark{a}} & \colhead{$E_{\rm {u}}/k$} \\
\colhead{} & \colhead{} & \colhead{(GHz)} & \colhead{(K)}
}
\startdata
HC$_{3}$N & 9$-$8 & 81.88147 & 19.6 \\
			   & 10$-$9 & 90.97902 & 24.0 \\
N$_{2}$H$^{+}$ & 1$-$0 & 93.17340 & 4.5 \\
 CCS & $6_{7}-5_{6}$ & 81.50517 & 15.4 \\
{\it {c}-}C$_{3}$H$_{2}$ & $2_{0,2} - 1_{1,1}$ & 82.09354 & 6.4 \\ 
\enddata
\tablenotetext{a}{Taken from the Cologne Database for Molecular Spectroscopy \citep[CDMS;][]{2005JMoSt.742..215M}.}
\end{deluxetable}	
%%%%%%%%%%%%%%%%%%%%%%%%%%%%%%%%%%%%%%%%%%%%%%%%%%%%%%%%%%%%%%%%%%%%                

%% Note that the \setcounter and \renewcommand are needed here because
%% this example is using a mix of deluxetable and tabular.  Here the
%% deluxetable counters are set with \tablenum but the situation is a bit
%% more complex for tabular.  Use the first command to set the Table number
%% to ONE LESS than it should be.  The next command will auto increment it
%% to the desired number.

\section{Results} \label{sec:res}

We conducted data reduction with the Java Newstar, the astronomical data analyzing system of the Nobeyama 45-m telescope data.
We obtained spectra with almost uniform rms noise levels ($\sim 15$ mK), which are comparable to those in the 45 GHz band results \citep[$\sim 10$ mK;][]{2018ApJ...854..133T}.
Our results are the deepest survey observations of carbon-chain molecules in high-mass star-forming regions \citep[e.g., $\approx 70-100$ mK;][]{2008ApJ...678.1049S}.
We set the criteria for line detection as a signal-to-noise (S/N) ratio above 4, and for tentative detection as an S/N ratio above 3.

\begin{figure}
\figurenum{1}
\plotone{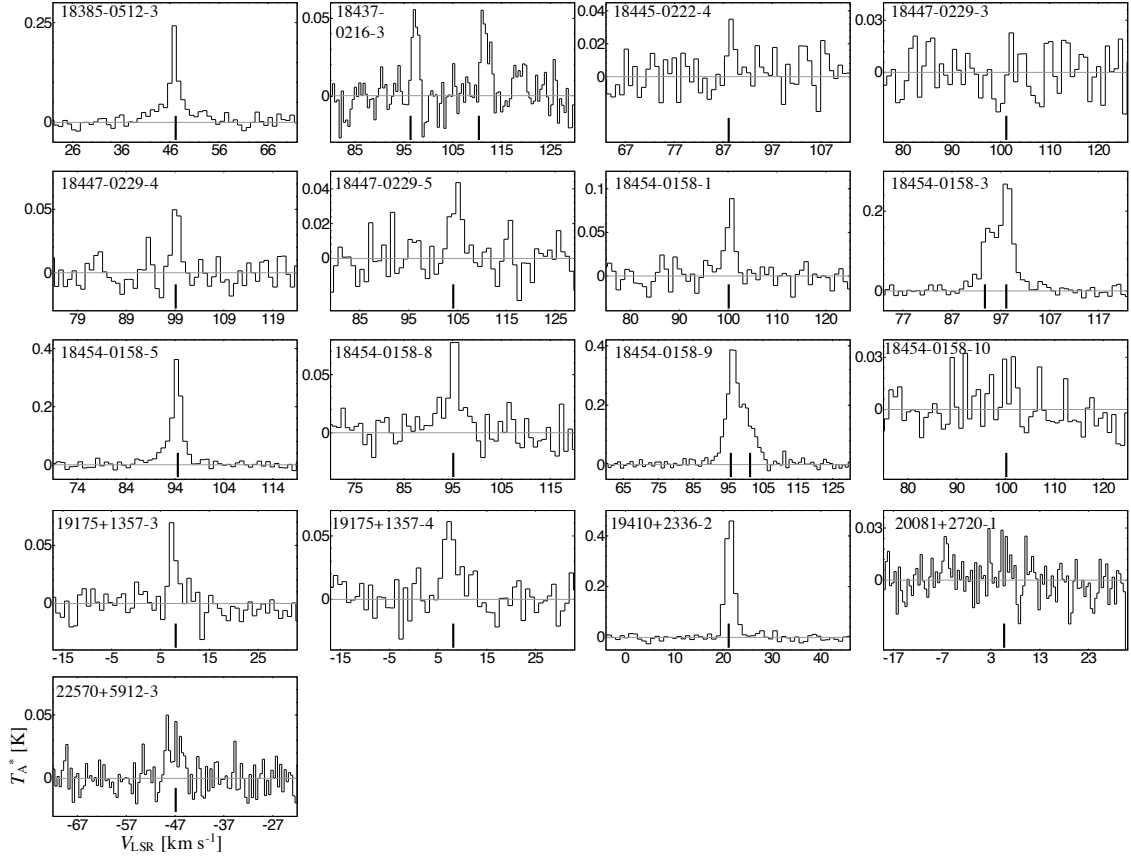}
\caption{Spectra of HC$_{3}$N ($J=9-8$) in HMSCs. The vertical lines indicate the systemic velocities of each source reported by \citep{2005ApJ...634L..57S}, except for HMSC 18385-0512-3. In the case of HMSC 18385-0512-3, the vertical line corresponds to the $V_{\rm {LSR}}$ value of the $J=5-4$ transition line of HC$_{3}$N reported by \citep{2018ApJ...854..133T}. \label{fig:f1}}
\end{figure}

\begin{figure}
\figurenum{2}
\plotone{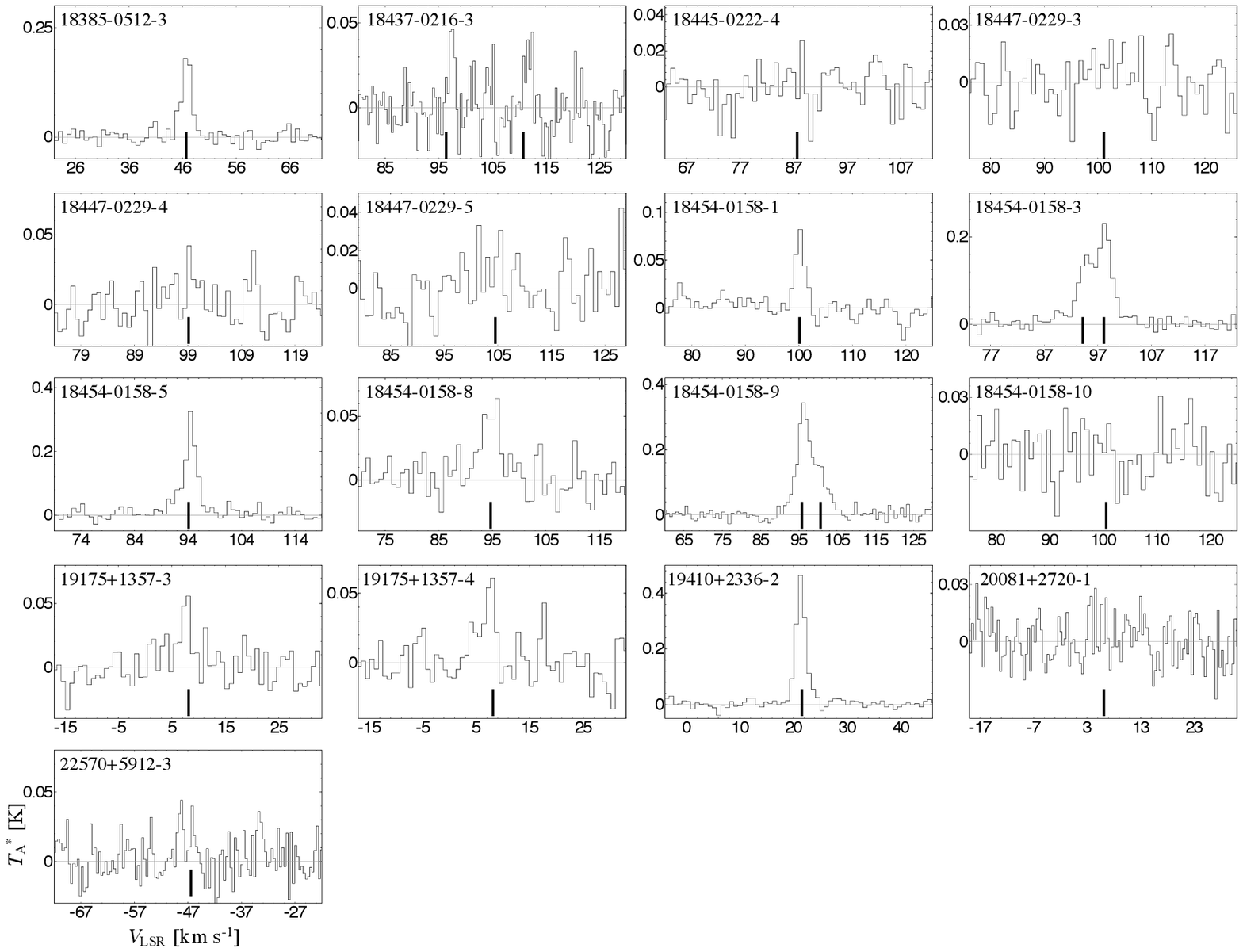}
\caption{Spectra of HC$_{3}$N ($J=10-9$) in HMSCs. The vertical lines indicate the systemic velocities of each source reported by \citep{2005ApJ...634L..57S}, except for HMSC 18385-0512-3. In the case of HMSC 18385-0512-3, the vertical line corresponds to the $V_{\rm {LSR}}$ value of the $J=5-4$ transition line of HC$_{3}$N reported by \citep{2018ApJ...854..133T}. \label{fig:f2}}
\end{figure}

\begin{figure}
\figurenum{3}
\plotone{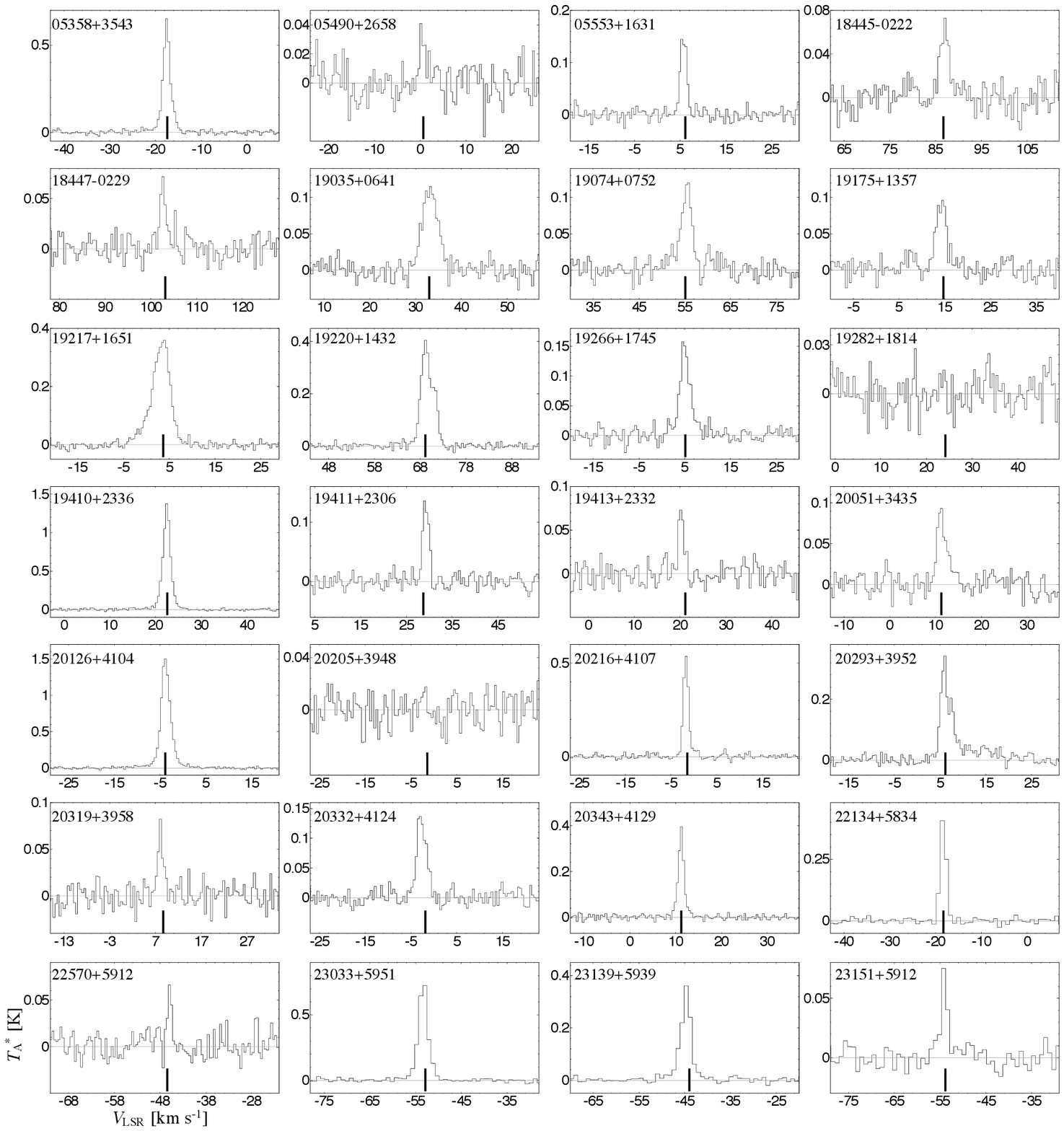}
\caption{Spectra of HC$_{3}$N ($J=9-8$) in HMPOs. The vertical lines indicate the systemic velocities of each source reported by \citep{2002ApJ...566..931S}. \label{fig:f3}}
\end{figure}

\begin{figure}
\figurenum{4}
\plotone{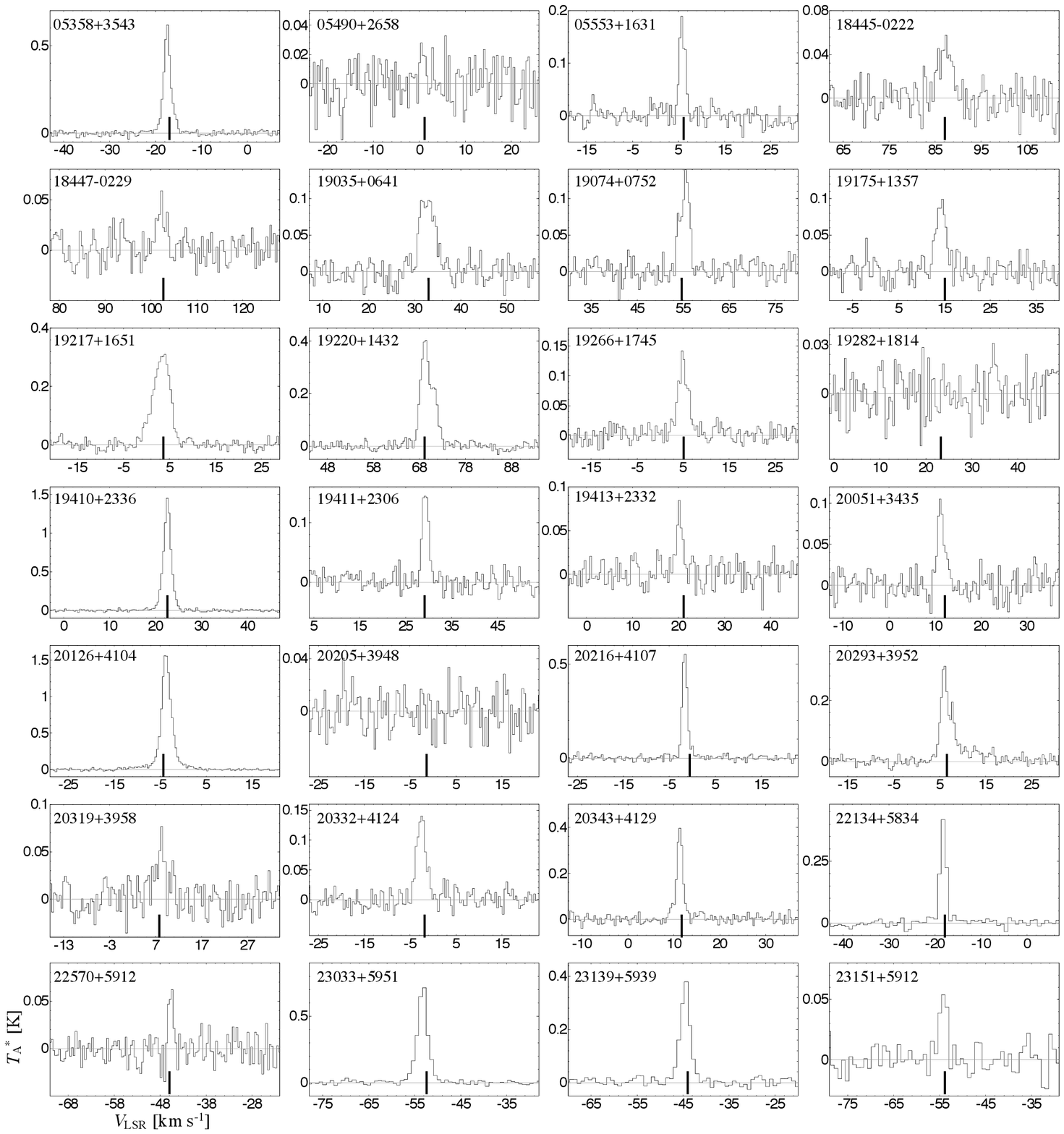}
\caption{Spectra of HC$_{3}$N ($J=10-9$) in HMPOs. The vertical lines indicate the systemic velocities of each source reported by \citep{2002ApJ...566..931S}. \label{fig:f4}}
\end{figure}

Figures \ref{fig:f1} -- \ref{fig:f4} show the spectra of HC$_{3}$N ($J=9-8$ and $10-9$) in HMSCs and HMPOs.
The HC$_{3}$N emission lines were detected in 14 HMSCs and in 26 HMPOs with an S/N ratio above 3, including tentative detection.
We fitted the spectra with a Gaussian profile to obtain spectral line parameters summarized in Table \ref{tab:tab2}.
In three HMSCs (18437-0216-3, 18454-0158-3, 18454-0158-9), the two velocity components are blended and we applied two-component Gaussian fitting.
The $V_{\rm {LSR}}$ values are consistent with the previous results \citep{2018ApJ...854..133T} within their errors.

%%%%%%%%%%%%%%%%%%%%%%%%%%%%%%%%%%%%%%%%%%%%%%%%%%%%%%%%%%%%%%%%%%

%%%%%%%%%%%%%%%%%%%%%%%%%%%%%%%%%%%%%%%%%%%%%%%%%%%%%%%%%%%%%%%%%%

%%%%%%%%%%%%%%%%%%%%%%%%%%%%%%%%%%%%%%%%%%%%%%%%%%%%%%%%%%%%%%%%%%%%
\floattable
\begin{deluxetable}{lccccccccccc}
\tabletypesize{\scriptsize}
\tablecaption{Spectral line parameters of HC$_{3}$N in HMSCs and HMPOs \label{tab:tab2}}
\tablewidth{0pt}
\tablehead{
\colhead{} & \multicolumn{5}{c}{HC$_{3}$N ($J=9-8$)} & \multicolumn{5}{c}{HC$_{3}$N ($J=10-9$)} \\
\cline{2-6}\cline{8-12}
\colhead{Source} & \colhead{{\it T}$^{\ast}_{\mathrm A}$} & \colhead{$\Delta v$\tablenotemark{a}} & \colhead{{\it V}$_{\mathrm {LSR}}$\tablenotemark{b}} & \colhead{$\int T^{\ast}_{\mathrm A}dv$} & \colhead{rms\tablenotemark{c}} & \colhead{} & \colhead{{\it T}$^{\ast}_{\mathrm A}$} & \colhead{$\Delta v$\tablenotemark{a}} & \colhead{{\it V}$_{\mathrm {LSR}}$\tablenotemark{b}} & \colhead{$\int T^{\ast}_{\mathrm A}dv$} & \colhead{rms\tablenotemark{c}} \\
\colhead{} & \colhead{(K)} & \colhead{(km s$^{-1}$)} & \colhead{(km s$^{-1}$)} & \colhead{(K km s$^{-1}$)} & \colhead{(mK)} &\colhead{} & \colhead{(K)} & \colhead{(km s$^{-1}$)} & \colhead{(km s$^{-1}$)} & \colhead{(K km s$^{-1}$)} & \colhead{(mK)}
}
\startdata
\multicolumn{11}{l}{HMSCs}  \\
	18385-0512-3 & 0.226 (15) & 1.87 (14) & 46.8 & 0.45 (4) & 10.9 & & 0.174 (13) & 2.4 (2) & 46.5 & 0.44 (5) & 13.6 \\
	18437-0216-3 & 0.045 (8) & 2.2 (5) & 111.0 & 0.11 (3) & 10.4 & & $<0.045$ & ... & ... & ... & 15.0 \\
			      & 0.056 (10) & 1.4 (3) & 97.2 & 0.09 (2) & 10.4 & & ... & ... & ... & ... & 15.0 \\
	18445-0222-4 & 0.035 (10) & 1.5 (5) & 88.6 & 0.06 (2) & 10.4 & & $<0.036$ & ... & ... & ... & 11.9 \\
	18447-0229-3 & $<0.037$  & ... & ... & ... & 12.3 & & $<0.037$ & ... & ... & ... & 12.4 \\
	18447-0229-4 & 0.050 (10) & 2.0 (5) & 98.9 & 0.11 (3) & 10.4 & & $<0.047$ & ... & ... & ... & 15.5 \\
	18447-0229-5 & 0.038 (8) & 2.8 (7) & 105.1 & 0.11 (4) & 10.4 & & $<0.043$ & ... & ... & ... & 14.4 \\
	18454-0158-1 & 0.085 (10) & 2.0 (3) & 100.8 & 0.18 (3) & 10.4 & & 0.079 (9) & 2.1 (3) & 100.3 & 0.18 (3) & 11.0 \\
	18454-0158-3 & 0.274 (11) & 2.61 (12) & 97.9 & 0.76 (5) & 10.1 & & 0.214 (10) & 2.92 (16) & 98.2 & 0.66 (5) & 12.0 \\
			       & 0.160 (9) & 3.3 (3) & 94.3 & 0.57 (5) &10.1 & & 0.149 (9) & 3.5 (3) & 95.0 & 0.55 (5) & 12.0 \\
	18454-0158-5 & 0.330 (13) & 2.43 (11) &94.3 & 0.85 (5) & 9.6 & & 0.284 (14) & 2.72 (15) & 94.4 & 0.82 (6) & 15.2 \\
	18454-0158-8 & 0.066 (9) & 3.4 (5) & 95.8 & 0.24 (5) & 11.5 & & 0.055 (8) & 4.4 (8) & 96.0 & 0.26 (6) & 12.8 \\
	18454-0158-9 & 0.162 (10) & 5.6 (5) & 99.5 & 0.96 (10) & 10.8 & & 0.180 (9) & 8.2 (3) & 98.6 & 1.57 (10) & 14.3 \\
			       & 0.346 (13) & 3.70 (15) & 95.9 & 1.36 (8) & 10.8 & & 0.186 (13) & 2.29 (19) & 96.2 & 0.45 (5) & 14.3 \\
	18454-0158-10 & $<0.037$ & ... & ... & ... & 12.4 & & $<0.040$ & ... & ... & ... & 13.2 \\
	19175+1357-3 & 0.070 (11) & 1.5 (3) & 7.3 & 0.11 (3) & 10.8 & & 0.049 (11) & 2.3 (6) & 8.1 & 0.12 (4) & 13.2 \\
	19175+1357-4 & 0.060 (9) & 3.0 (5) & 7.3 & 0.19 (5) & 11.5 & & 0.061 (13) & 1.9 (5) & 8.1 & 0.12 (4) & 14.2 \\
	19410+2336-2 & 0.448 (19) & 2.27 (11) & 21.7 & 1.08 (7) & 11.1 & & 0.451 (11) & 2.19 (6) & 21.3 & 1.05 (4) & 13.3 \\
	20081+2720-1 & $<0.035$ & ... & ... & ... & 11.5 & & $<0.039$ & ... & ... & ... & 12.9 \\
	22570+5912-3 & 0.048 (11) & 1.0 (3) & -48.6 & 0.05 (2) & 10.9 & & $<0.042$ & ... & ... & ... & 13.9 \\ 
	& & & & & & & & & & \\
	\multicolumn{11}{l}{HMPOs}  \\
	05358+3543 & 0.602 (10) & 1.89 (4) & -17.6 & 1.21 (3) & 9.8 & & 0.573 (10) & 1.85 (4) & -17.4 & 1.13 (3) & 11.2 \\
	05490+2658 & 0.033 (8) & 1.4 (4) & 0.2 & 0.05 (2) & 10.4 & & $<0.041$ & ... & ... & ... &13.7 \\
	05553+1631 & 0.151 (8) & 1.70 (10) & 5.4 & 0.27 (2) & 9.8 & & 0.181 (10) & 1.51 (10) & 5.7 & 0.29 (3) & 14.1 \\
	18445-0222 & 0.064 (9) & 2.4 (3) & 87.0 & 0.16 (3) & 10.7 & & 0.048 (6) & 3.4 (5) & 87.4 & 0.17 (4) & 13.4 \\
	18447-0229 & 0.067 (9) & 1.5 (2) & 102.6 & 0.10 (2) & 11.0 & & 0.042 (8) & 2.3 (5) & 102.3 & 0.11 (3) & 13.8 \\
	19035+0641 & 0.113 (5) & 4.0 (2) & 33.3 & 0.48 (3) & 11.4 & & 0.099 (6) & 4.0 (3) & 33.1 & 0.42 (4) & 13.1 \\
	19074+0752 & 0.115 (7) & 2.7 (2) & 55.8 & 0.33 (3) & 11.5 & & 0.124 (8) & 2.44 (18) & 55.5 & 0.32 (3) & 13.6 \\
	19175+1357 & 0.091 (6) & 2.9 (2) & 14.5 & 0.28 (3) & 11.2 & & 0.092 (8) & 2.6 (3) & 14.6 & 0.26 (3) & 9.9 \\
	19217+1651 & 0.336 (10) & 4.61 (16) & 4.0 & 1.65 (8) & 10.7 & & 0.311 (7) & 4.21 (11) & 4.1 & 1.40 (5) & 13.7 \\
	19220+1432 & 0.360 (10) & 3.30 (10) & 69.2 & 1.26 (5) & 10.6 & & 0.365 (10) & 3.32 (10) & 69.5 & 1.29 (5) & 13.0 \\
	19266+1745 & 0.145 (6) & 2.66 (13) & 4.6 & 0.41 (3) & 9.5 & & 0.120 (6) & 2.74 (17) & 4.8 & 0.35 (3) & 11.9 \\
	19282+1814 & $<0.032$ & ... & ... & ... & 10.7 & & $<0.040$ & ... & ... & ... & 13.3 \\
	19410+2336 & 1.298 (12) & 1.97 (2) & 22.4 & 2.72 (4) & 11.3 & & 1.359 (13) & 1.96 (2) & 22.5 & 2.83 (4) & 13.3 \\
	19411+2306 & 0.130 (8) & 1.65 (12) & 29.0 & 0.23 (2) & 9.8 & & 0.159 (10) & 1.63 (11) & 29.2 & 0.28 (3) & 13.6 \\
	19413+2332 & 0.077 (9) & 1.32 (17) & 20.1 & 0.11 (2) & 10.7 & & 0.074 (10) & 1.5 (2) & 20.2 & 0.12 (2) & 13.5 \\
	20051+3435 & 0.085 (7) & 2.5 (2) & 11.3 & 0.23 (3) & 11.2 & & 0.098 (10) & 1.7 (2) & 11.1 & 0.18 (3) & 15.3 \\ 
	20126+4104 & 1.456 (14) & 2.40 (3) & -3.8 & 3.72 (6) & 12.2 & & 1.533 (17) & 2.49 (3) & -4.0 & 4.06 (7) & 14.0 \\
	20205+3948 & $<0.033$ & ... & ... & ... & 11.0 & & $<0.041$ & ... & ... & ... & 13.5 \\
	20216+4107 & 0.539 (10) & 1.33 (3) & -1.7 & 0.76 (2) & 11.7 & & 0.558 (13) & 1.36 (4) & -1.5 & 0.81 (3) & 13.8 \\
	20293+3952 & 0.281 (12) & 2.85 (14) & 6.0 & 0.85 (5) & 12.0 & & 0.265 (11) & 2.88 (13) & 6.2 & 0.81 (5) & 13.4 \\
	20319+3958 & 0.070 (9) & 1.5 (2) & 8.1 & 0.11 (2) & 12.0 & & 0.044 (7) & 3.6 (7) & 8.3 & 0.17 (4) & 15.2 \\
	20332+4124 & 0.133 (7) & 2.53 (16) & -3.0 & 0.36 (3) & 11.4 & & 0.138 (8) & 2.39 (16) & -2.5 & 0.35 (3) & 14.1 \\
	20343+4129 & 0.368 (9) & 1.47 (4) & 11.3 & 0.58 (2) & 11.0 & & 0.390 (13) & 1.53 (6) & 11.3 & 0.64 (3) & 16.4 \\
	22134+5834 & 0.409 (15) & 1.63 (7) & -18.6 & 0.71 (4) & 10.6 & & 0.421 (14) & 1.59 (6) & -18.2 & 0.71 (3) & 13.7 \\
	22570+5912 & 0.066 (10) & 1.02 (18) & -46.0 & 0.07 (2) & 11.3 & & 0.065 (11) & 1.1 (2) & -45.4 & 0.08 (2) & 13.4 \\
	23033+5951 & 0.69 (3) & 2.65 (11) & -53.0 & 1.95 (11) & 10.0 & & 0.67 (2) & 2.83 (12) & -52.9 & 2.00 (11) & 12.0 \\ 
	23139+5939 & 0.342 (9) & 2.58 (8) & -44.7 & 0.94 (4) & 8.1 & & 0.356 (13) & 2.60 (11) & -44.3 & 0.99 (5) & 12.5 \\
	23151+5912 & 0.074 (8) & 1.64 (19) & -54.3 & 0.13 (2) & 7.6 & & 0.059 (10) & 1.9 (4) & -54.2 & 0.12 (3) & 10.1 \\
\enddata
\tablecomments{The numbers in parentheses represent the standard deviation in the Gaussian fit. The errors are written in units of the last significant digit. The upper limits correspond to $3 \sigma$ values.}
\tablenotetext{a}{These values are not corrected for instrumental velocity resolution.}
\tablenotetext{b}{The error of $V_{\rm {LSR}}$ is commonly 0.8 km s$^{-1}$, corresponding to the velocity resolution of the final spectra (Section \ref{sec:obs}).}
\tablenotetext{c}{The rms noises were evaluated in emission-free region.}
\end{deluxetable}	
%%%%%%%%%%%%%%%%%%%%%%%%%%%%%%%%%%%%%%%%%%%%%%%%%%%%%%%%%%%%%%%%%%

\begin{figure}
\figurenum{5}
\plotone{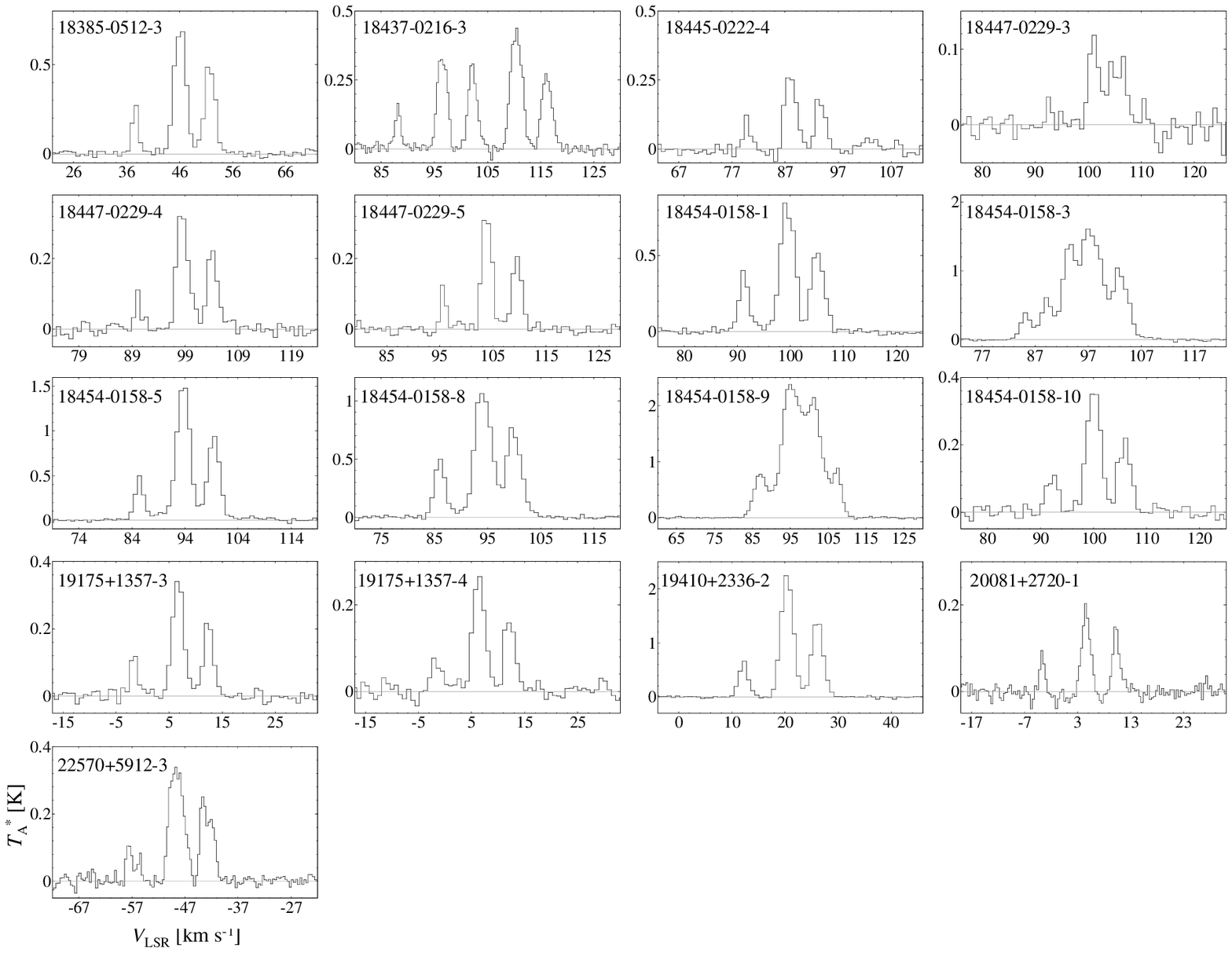}
\caption{Spectra of N$_{2}$H$^{+}$ ($J=1-0$) in HMSCs.\label{fig:f5}}
\end{figure}

\begin{figure}
\figurenum{6}
\plotone{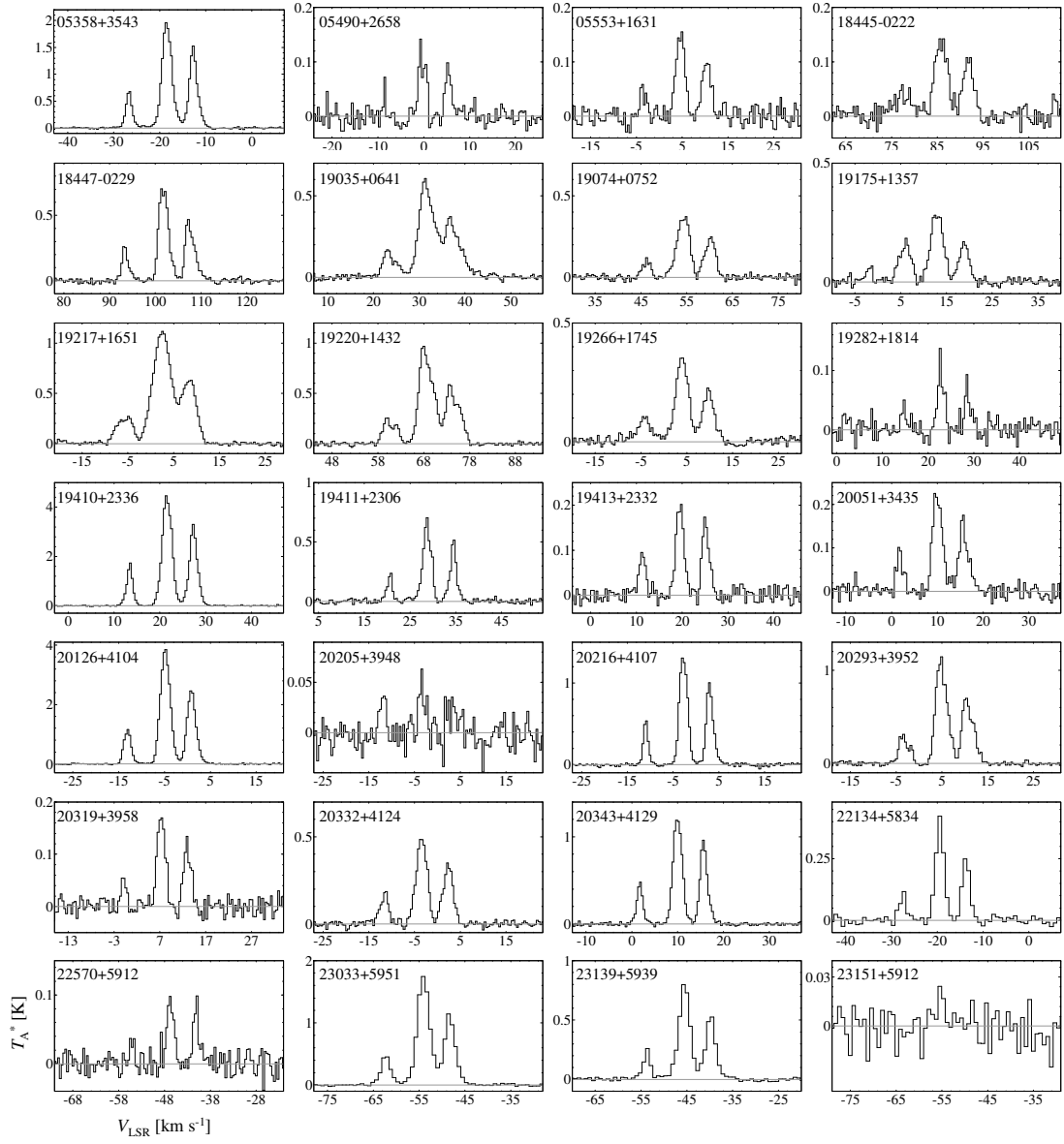}
\caption{Spectra of N$_{2}$H$^{+}$ ($J=1-0$) in HMPOs.\label{fig:f6}}
\end{figure}

Figures \ref{fig:f5} and \ref{fig:f6} show the spectra of N$_{2}$H$^{+}$ ($J=1-0$) in HMSCs and HMPOs, respectively.
Table \ref{tab:t3} summarizes the spectral line parameters.
We set the $F_{1} = 2-1$ transition as the criteria of velocity, and the $V_{\rm {LSR}}$ values are consistent with those of HC$_{3}$N within their errors.
N$_{2}$H$^{+}$ was detected in all of the target sources except for HMPO 23151+5912 with an S/N ratio above 4.
Seven hyperfine components were detected as three lines.
In some sources, we cannot clearly identify lines due to blending two velocity components.

%%%%%%%%%%%%%%%%%%%%%%%%%%%%%%%%%%%%%%%%%%%%%%%%%%%%%%%%%%%%%%%%%%

%%%%%%%%%%%%%%%%%%%%%%%%%%%%%%%%%%%%%%%%%%%%%%%%%%%%%%%%%%%%%%%%%
\clearpage
%clearpage is must for longtable
%%%%%%%%%%%%%%%%%%%%%%%%%%%%%%%%%%%%%%%%%%%%%%%%%%%%%%%%%%%%%%%%%%%%%%%%%
\startlongtable
\begin{deluxetable*}{lcccccc}
\tablecaption{Spectral line parameters of N$_{2}$H$^{+}$ ($J=1-0$) in HMSCs and HMPOs \label{tab:t3}}
\tablewidth{700pt}
\tabletypesize{\scriptsize}
\tablehead{
\colhead{Source} & \colhead{Hyperfine} & \colhead{{\it T}$^{\ast}_{\rm A}$} & \colhead{$\Delta v$\tablenotemark{a}} & \colhead{{\it V}$_{\rm LSR}$\tablenotemark{b}} & \colhead{$\int T^{\ast}_{\rm A}dv$} & \colhead{rms\tablenotemark{c}} \\
\colhead{ } & \colhead{transition} & \colhead{(K)} & \colhead{(km/s)} & \colhead{(km/s)} & \colhead{(K km/s)} & \colhead{(mK)}
} 
\startdata
         \multicolumn{7}{l}{HMSCs} \\
	18385-0512-3 & $F_{1} = 1-1$ & 0.507 (15) & 2.34 (8) & 51.7 & 1.26 (6) & 14.7 \\
			       & $F_{1} = 2-1$ & 0.724 (14) & 2.63 (6) & 46.0 & 2.03 (6) &   \\
			       & $F_{1} = 0-0$ & 0.29 (2) & 1.33 (11) & 37.6 & 0.41 (5) & \\
	18437-0216-3\tablenotemark{d} & 			& 0.269 (9) & 2.27 (9) & 116.1 & 0.65 (3) & 13.0 \\
			       &			& 0.436 (9) & 2.52 (6) & 110.4 & 1.17 (4) &   \\
			       & 			& 0.308 (10) & 1.92 (7) & 102.2 & 0.63 (3) & \\
			       &			& 0.350 (10) & 2.16 (7) & 96.5 & 0.80 (3) & \\
			       &			& 0.156 (13) & 1.21 (11) & 88.1 & 0.20 (3) & \\
	18445-0222-4 & $F_{1} = 1-1$ & 0.185 (15) &  2.2 (2) & 93.6 & 0.43 (5) & 14.2 \\
			      & $F_{1} = 2-1$ & 0.279 (14) & 2.47 (14) & 88.1 & 0.73 (6) & \\
			      & $F_{1} = 0-0$ & 0.122 (18) & 1.3 (2) & 79.8 & 0.17 (3) & \\
	18447-0229-3\tablenotemark{d} & $F_{1} = 1-1$ ? & 0.078 (9) & 4.4 (7) & 105.4 & 0.37 (7) & 13.8 \\
			      & $F_{1} = 2-1$ ? & 0.116 (13) & 2.0 (3) & 100.9 & 0.24 (4) & \\
	18447-0229-4 & $F_{1} = 1-1$ & 0.226 (12) & 2.34 (14) & 104.1 & 0.56 (5) & 13.3 \\      
			       & $F_{1} = 2-1$ & 0.334 (11) & 2.59 (10) & 98.4 & 0.92 (5) & \\
			       & $F_{1} = 0-0$ & 0.110 (15) & 1.15 (16) & 90.1 & 0.13 (3) & \\
	18447-0229-5 & $F_{1} = 1-1$ & 0.209 (11) & 2.14 (12) & 109.6 & 0.48 (3) & 11.0 \\    
			       & $F_{1} = 2-1$ & 0.337 (10) & 2.17 (7) & 103.8 & 0.78 (3) & \\
			       & $F_{1} = 0-0$ & 0.133 (13) & 1.19 (13) &  95.7 & 0.17 (2) & \\
	18454-0158-1 & $F_{1} = 1-1$ & 0.539 (17) & 2.82 (11) & 105.1 & 1.62 (8) & 13.2 \\
			       & $F_{1} = 2-1$ & 0.858 (16) & 3.15 (7) & 99.4 & 2.87 (8) & \\ 
			       & $F_{1} = 0-0$ & 0.39 (2) & 2.09 (12) &  91.2 & 0.87 (7) & \\
	18454-0158-3\tablenotemark{d} & 			& 0.94 (2) & 3.35 (10) & 103.1 & 3.36 (13) & 13.4 \\
			       &			& 1.610 (19) & 4.92 (10) & 97.4 & 8.4 (2) & \\
			       &			& 1.16 (3) & 2.47 (7) & 93.2 & 3.07 (12) & \\
			       &			& 0.58 (3) & 2.62 (15) & 89.3 & 1.63 (12) & \\
			       &			& 0.38 (3) & 2.3 (2) & 85.3 & 0.94 (10) & \\ 
	18454-0158-5 & $F_{1} = 1-1$ & 0.946 (17) & 2.62 (5) & 99.4 & 2.64 (7) & 14.7 \\    			        
			       & $F_{1} = 2-1$ & 1.548 (16) & 2.94 (3) & 93.7 & 4.83 (8) & \\
			       & $F_{1} = 0-0$ & 0.477 (19) & 1.96 (9) & 85.4 & 0.99 (6) & \\
	18454-0158-8 & $F_{1} = 1-1$ & 0.749 (16) & 3.30 (9) & 99.9 & 2.63 (9) & 12.1 \\    
			       & $F_{1} = 2-1$ & 1.108 (16) & 3.63 (6) & 94.2 & 4.28 (9) & \\
			       & $F_{1} = 0-0$ & 0.502 (19) & 2.40 (10) & 86.0 & 1.28 (7) & \\
	18454-0158-9\tablenotemark{d} & 			& 0.79 (3) & 2.94 (13) & 107.3 & 2.48 (14) & 11.0 \\
                                & 			& 2.00 (2) & 5.62 (9) & 101.0 & 12.0 (2) & \\
                                &			& 2.34 (2) & 5.28 (7) & 95.0 & 13.1 (2) & \\
                                &			& 0.78 (2) & 4.17 (16) & 87.2 & 3.48 (17) & \\ 
       18454-0158-10 & $F_{1} = 1-1$ & 0.210 (11) & 2.82 (18) & 105.8 & 0.63 (5) & 13.9 \\    
       			       & $F_{1} = 2-1$ & 0.373 (11) & 3.07 (10) & 100.0 & 1.22 (5) & \\
			       & $F_{1} = 0-0$ & 0.108 (12) & 2.6 (3) & 92.2 & 0.30 (5) & \\
	19175+1357-3 & $F_{1} = 1-1$ & 0.225 (10) & 2.14 (11) & 12.4 & 0.51 (4) & 11.7 \\ 
			        & $F_{1} = 2-1$ & 0.351 (10) & 2.52 (8) & 6.7 & 0.94 (4) & \\
			        & $F_{1} = 0-0$ & 0.133 (12) & 1.57 (17) & -1.6 & 0.22 (3) & \\
	19175+1357-4 & $F_{1} = 1-1$ & 0.172 (11) & 2.39 (18) & 12.1 & 0.44 (4) & 14.3 \\
			        & $F_{1} = 2-1$ & 0.269 (11) & 2.72 (12) & 6.4 & 0.78 (5) & \\
			        & $F_{1} = 0-0$ & 0.070 (11) & 2.5 (4) & -1.6 & 0.19 (5) & \\
	19410+2336-2 & $F_{1} = 1-1$ & 1.422 (16) & 2.51 (3) & 26.1 & 3.80 (6) & 13.7 \\
			        & $F_{1} = 2-1$ & 2.274 (15) & 2.71 (2) & 20.4 & 6.55 (7) & \\
			        & $F_{1} = 0-0$ & 0.646 (17) & 2.04 (6) & 12.1 & 1.40 (6) & \\ 
	20081+2720-1 & $F_{1} = 1-1$ & 0.149 (9) & 1.46 (11) & 10.2 & 0.23 (2) & 12.0 \\
				& $F_{1} = 2-1$ & 0.191 (8) & 1.85 (9) & 4.6 & 0.38 (2) & \\
				& $F_{1} = 0-0$ & 0.094 (11) & 0.97 (14) & -3.4 & 0.10 (2) & \\
	22570+5912-3 & $F_{1} = 1-1$ & 0.228 (11) & 2.81 (16) & -42.9 & 0.68 (5) & 13.5 \\
				& $F_{1} = 2-1$ & 0.354 (11) & 3.02 (11) & -48.6 & 1.14 (5) & \\
				& $F_{1} = 0-0$ & 0.084 (12) & 2.5 (4) & -57.1 & 0.22 (5) & \\
	 & & & & & & \\		
         \multicolumn{7}{l}{HMPOs} \\
	05358+3543 & $F_{1} = 1-1$ & 1.42 (2) & 2.08 (3) & -12.8 & 3.14 (7) & 10.7 \\
			    & $F_{1} = 2-1$ & 1.950 (18) & 2.51 (3) & -18.4 & 5.21 (7) & \\
			    & $F_{1} = 0-0$ & 0.67 (2) & 1.59 (6) & -27.0 & 1.13 (6) & \\
	05490+2658 & $F_{1} = 1-1$ & 0.092 (17) & 1.5 (3) & 5.3 & 0.14 (4) & 14.8 \\
			    & $F_{1} = 2-1$ & 0.107 (13) & 2.0 (3) & -0.3 & 0.23 (4) & \\
			    & $F_{1} = 0-0$ & 0.07 (3) & 0.4 (2) & -8.7 & 0.03 (2) & \\
	05553+1631 & $F_{1} = 1-1$ & 0.102 (8) & 2.2 (2) & 10.3 & 0.24 (3) & 13.8 \\
			    & $F_{1} = 2-1$ & 0.156 (8) & 2.22 (13) & 4.6 & 0.37 (3) & \\
			    & $F_{1} = 0-0$ & 0.050 (9) & 1.7 (4) & -3.6 & 0.09 (2) & \\ 
	18445-0222 & $F_{1} = 1-1$ & 0.106 (9) & 2.7 (3) & 91.8 & 0.30 (4) & 13.7 \\
			    & $F_{1} = 2-1$ & 0.140 (8) & 3.3 (2) & 85.8 & 0.50 (4) & \\
	18447-0229  & $F_{1} = 1-1$ & 0.430 (13) & 2.45 (9) & 107.5 & 1.12 (5) & 13.7 \\
			    & $F_{1} = 2-1$ & 0.727 (13) & 2.42 (5) & 101.8 & 1.87 (5) & \\
			    & $F_{1} = 0-0$ & 0.253 (16) & 1.54 (12) & 93.4 & 0.41 (4) & \\
	19035+0641 & $F_{1} = 1-1$ & 0.316 (7) & 5.36 (16) & 36.8 & 1.80 (7) & 11.9 \\
			    & $F_{1} = 2-1$ & 0.566 (9) & 3.40 (6) & 31.4 & 2.05 (5) & \\
			    & $F_{1} = 0-0$ & 0.148 (8) & 3.6 (2) & 23.5 & 0.57 (5) & \\
	19074+0752 & $F_{1} = 1-1$ & 0.246 (9) & 2.62 (11) & 60.2 & 0.68 (4) & 13.6 \\
			    & $F_{1} = 2-1$ & 0.381 (8) & 3.17 (8) & 54.3 & 1.28 (4) & \\
			    & $F_{1} = 0-0$ & 0.102 (10) & 2.1 (2) & 46.2 & 0.23 (3) & \\
	19175+1357 & $F_{1} = 1-1$ & 0.164 (9) & 2.50 (16) & 18.9 & 0.44 (40 & 12.8 \\
			    & $F_{1} = 2-1$ & 0.293 (8) & 3.19 (10) & 12.9 & 0.99 (4) & \\
			    & $F_{1} = 0-0$ & 0.162 (8) & 3.00 (18) & 6.0 & 0.52 (4) & \\ 
	19217+1651 & $F_{1} = 1-1$ & 0.640 (8) & 3.58 (6) & 8.4 & 2.44 (5) & 13.8 \\
			    & $F_{1} = 2-1$ & 1.117 (7) & 4.57 (4) & 2.5 & 5.43 (6) & \\
			    & $F_{1} = 0-0$ & 0.264 (8) & 4.07 (14) & -5.7 & 1.14 (5) & \\
	19220+1432\tablenotemark{d} & 			     & 0.503 (15) & 4.15 (15) & 74.2 & 2.22 (10) & 13.1 \\
			    & 		 	     & 0.936 (16) & 3.41 (7) & 68.4 & 3.39 (9) & \\
			    &			     & 0.18 (3) & 1.4 (2) & 62.1 & 0.28 (6) & \\
			    &			     & 0.25 (2) & 2.0 (2) & 59.9 & 0.53 (7) & \\
	19266+1745 & $F_{1} = 1-1$ & 0.212 (7) & 2.91 (11) & 9.7 & 0.66 (3) & 12.2 \\
			    & $F_{1} = 2-1$ & 0.356 (7) & 3.18 (7) & 4.0 & 1.20 (3) & \\
			    & $F_{1} = 0-0$ & 0.090 (6) & 4.0 (3) & -4.2 & 0.39 (4) & \\	
	19282+1814 & $F_{1} = 1-1$ & 0.083 (12) & 1.12 (19) & 28.5 & 0.10 (2) & 13.6 \\
			    & $F_{1} = 2-1$ & 0.107 (10) & 1.63 (17) & 22.7 & 0.19 (3) & \\
			    & $F_{1} = 0-0$ & 0.043 (12) & 1.1 (4) & 14.6 & 0.05 (2) & \\
	19410+2336 & $F_{1} = 1-1$ & 3.10 (4) & 2.04 (3) & 27.3 & 6.73 (12) & 14.5 \\
			    & $F_{1} = 2-1$ & 4.44 (3) & 2.47 (2) & 21.6 & 11.69 (13) & \\
			    & $F_{1} = 0-0$ & 1.64 (4) & 1.50 (4) & 13.4 & 2.61 (10) & \\
	19411+2306 & $F_{1} = 1-1$ & 0.489 (16) & 1.57 (6) & 34.5 & 0.82 (4) & 16.3 \\
			    & $F_{1} = 2-1$ & 0.651 (13) & 2.16 (5) & 28.7 & 1.50 (5) & \\
			    & $F_{1} = 0-0$ & 0.224 (17) & 1.34 (12) & 20.6 & 0.32 (4) & \\
	19413+2332 & $F_{1} = 1-1$ & 0.166 (9) & 1.78 (11) & 25.1 & 0.31 (3) & 13.1 \\
			    & $F_{1} = 2-1$ & 0.204 (8) & 2.03 (9) & 19.5 & 0.44 (3) & \\
			    & $F_{1} = 0-0$ & 0.094 (9) & 1.65 (18) & 11.3 & 0.17 (2) & \\
	20051+3435 & $F_{1} = 1-1$ & 0.146 (8)& 2.53 (16) & 15.7 & 0.39 (3) & 14.7 \\
			    & $F_{1} = 2-1$ & 0.226 (8) & 2.50 (11) & 10.0 & 0.60 (3) & \\
			    & $F_{1} = 0-0$ & 0.090 (10) & 1.7 (2) & 1.8 & 0.16 (3) & \\
	20126+4104 & $F_{1} = 1-1$ & 2.431 (18) & 2.25 (2) & 0.9 & 5.82 (6) & 12.9 \\
			    & $F_{1} = 2-1$ & 3.855 (17) & 2.49 (1) & -4.8 & 10.20 (7) & \\
			    & $F_{1} = 0-0$ & 1.15 (2) & 1.83 (4) & -13.1 & 2.23 (6) & \\
	20205+3948 & $F_{1} = 2-1$ & 0.050 (10) & 1.6 (4) & -3.4 & 0.08 (2) & 14.3 \\
			    & $F_{1} = 0-0$ & 0.039 (10) & 1.5 (4) & -11.9 & 0.06 (2) & \\
	20216+4107 & $F_{1} = 1-1$ & 0.973 (16) & 1.77 (3) & 3.0 & 1.83 (5) & 14.2 \\
			     & $F_{1} = 2-1$ & 1.353 (15) & 2.09 (3) & -2.7 & 3.00 (5) & \\
			     & $F_{1} = 0-0$ & 0.55 (2) & 1.19 (5) & -11.0 & 0.70 (4) & \\		     
	20293+3952 & $F_{1} = 1-1$ & 0.657 (19) & 3.02 (10) & 10.6 & 2.11 (9) & 15.3 \\
			    & $F_{1} = 2-1$ & 1.094 (19) & 2.90 (6) & 4.8 & 3.37 (9) & \\
			    & $F_{1} = 0-0$ & 0.28 (2) & 2.8 (2) & -3.3 & 0.81 (9) & \\
	20319+3958 & $F_{1} = 1-1$ & 0.118 (9) & 1.88 (17) & 12.9 & 0.24 (3) & 13.6 \\
			    & $F_{1} = 2-1$  & 0.177 (9) & 2.05 (12) & 7.3 & 0.39 (3) & \\
			    & $F_{1} = 0-0$ & 0.058 (11) & 1.2 (3) & -1.0 & 0.08 (2) & \\
	20332+4124 & $F_{1} = 1-1$ & 0.335 (9) & 2.58 (8) & 2.3 & 0.92 (4) & 13.4 \\
			    & $F_{1} = 2-1$ & 0.496 (8) & 2.76 (5) & -3.4 & 1.45 (4) & \\
			    & $F_{1} = 0-0$ & 0.172 (9) & 2.15 (14) & -11.6 & 0.39 (3) & \\
	20343+4129 & $F_{1} = 1-1$ & 0.936 (14) & 1.88 (3) & 15.7 & 1.87 (4) & 13.5 \\
			    & $F_{1} = 2-1$ & 1.236 (12) & 2.28 (3) & 10.0 & 3.00 (5) & \\
			    & $F_{1} = 0-0$ & 0.468 (15) & 1.51 (6) & 1.7 & 0.75 (4) & \\
	22134+5834 & $F_{1} = 1-1$ & 0.259 (11) & 1.94 (10) & -13.7 & 0.53 (3) & 11.9 \\
			    & $F_{1} = 2-1$ & 0.413 (11) & 2.12 (6) & -19.4 & 0.93 (4) & \\
			    & $F_{1} = 0-0$ & 0.111 (12) & 1.7 (2) & -27.5 & 0.20 (3) & \\
	22570+5912 & $F_{1} = 1-1$ & 0.088 (11) & 1.31 (19) & -41.1 & 0.12 (2) & 13.7 \\
			    & $F_{1} = 2-1$ & 0.094 (9) & 1.8 (2) & -46.7 & 0.18 (3) & \\
			    & $F_{1} = 0-0$ & 0.037 (12) & 1.0 (4) & -55.1 & 0.04 (2) & \\
	23033+5951 & $F_{1} = 1-1$ & 1.097 (17) & 2.94 (5) & -48.4 & 3.43 (8) & 13.5 \\
			    & $F_{1} = 2-1$ & 1.745 (16) & 3.13 (3) & -54.1 & 5.81 (8) & \\
			    & $F_{1} = 0-0$ & 0.465 (18) & 2.53 (11) & -62.3 & 1.25 (7) & \\
	23139+5939 & $F_{1} = 1-1$ & 0.530 (13) & 2.58 (7) & -39.8 & 1.45 (5) & 11.0 \\
			    & $F_{1} = 2-1$ & 0.796 (12) & 2.89 (5) & -45.5 & 2.45 (6) & \\
			    & $F_{1} = 0-0$ & 0.252 (15) & 1.74 (12) & -53.7 & 0.47 (4) & \\
	23151+5912 & 			     & $< 0.030$ & ... & ... & ... & 9.6 \\  
\enddata
\tablecomments{The numbers in parentheses represent the standard deviation in the Gaussian fit. The errors are written in units of the last significant digit. The upper limits correspond to $3 \sigma$ values.}
\tablenotetext{a}{These values are not corrected for instrumental velocity resolution.}
\tablenotetext{b}{The error of $V_{\rm {LSR}}$ is commonly 0.8 km s$^{-1}$, corresponding to the velocity resolution of the final spectra (Section \ref{sec:obs}).}
\tablenotetext{c}{The rms noises were evaluated in emission-free region.}
\tablenotetext{d}{The hyperfine transition cannot be identified due to several velocity components are blended or broad line widths.}
\end{deluxetable*}
%%%%%%%%%%%%%%%%%%%%%%%%%%%%%%%%%%%%%%%%%%%%%%%%%%%%%%%%%%%%%%%%%

\begin{figure}
\figurenum{7}
\plotone{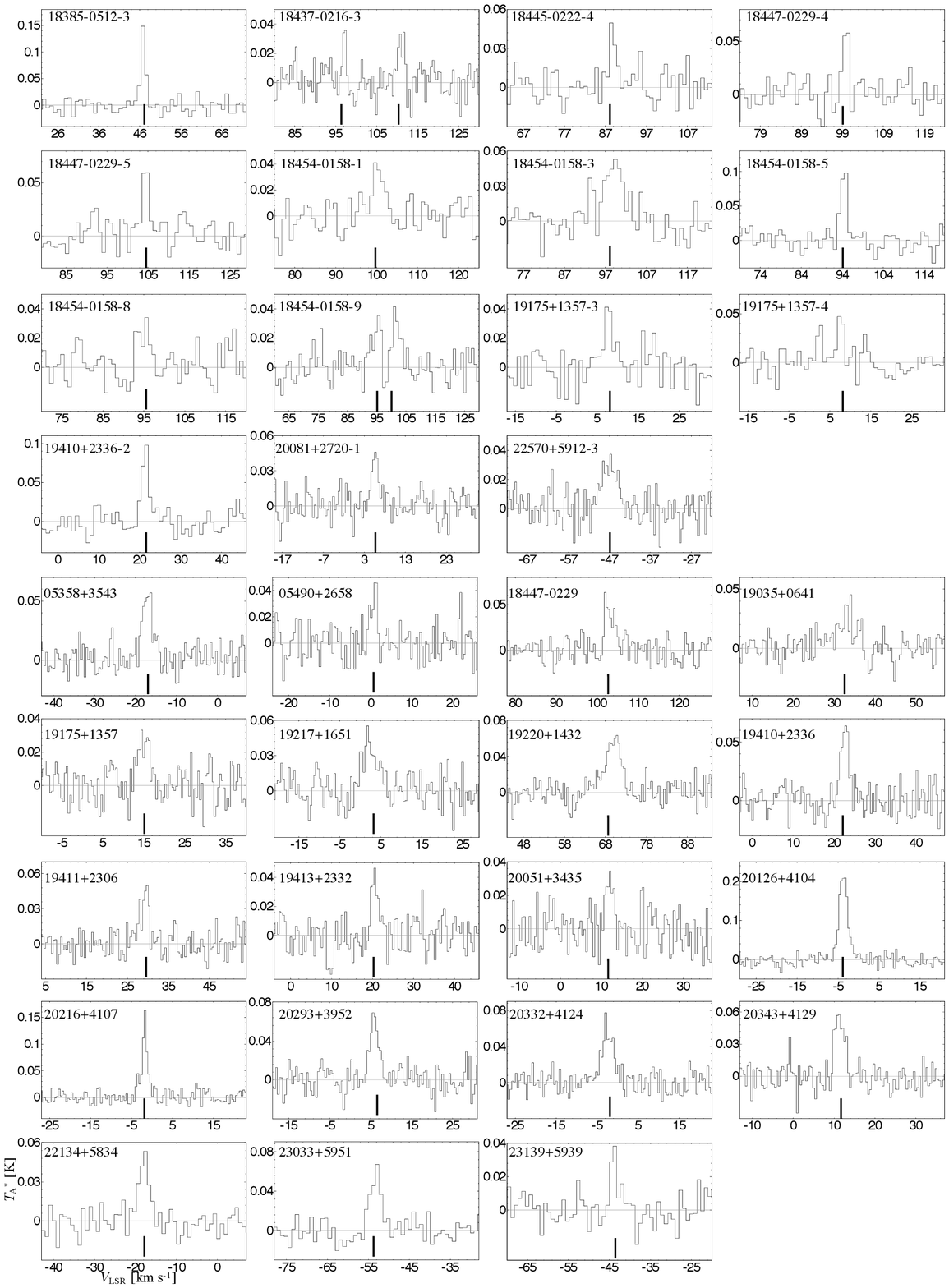}
\caption{Spectra of {\it c}-C$_{3}$H$_{2}$ ($J_{\rm Ka, Kc} =2_{0,2} - 1_{1,1}$) in HMSCs and HMPOs. The vertical lines indicate the systemic velocities of each source reported by \citep{2002ApJ...566..931S,2005ApJ...634L..57S}, except for HMSC 18385-0512-3. In the case of HMSC 18385-0512-3, the vertical line corresponds to the $V_{\rm {LSR}}$ value of the $J=5-4$ transition line of HC$_{3}$N reported by \citep{2018ApJ...854..133T}. \label{fig:f7}}
\end{figure}

\begin{figure}
\figurenum{8}
\plotone{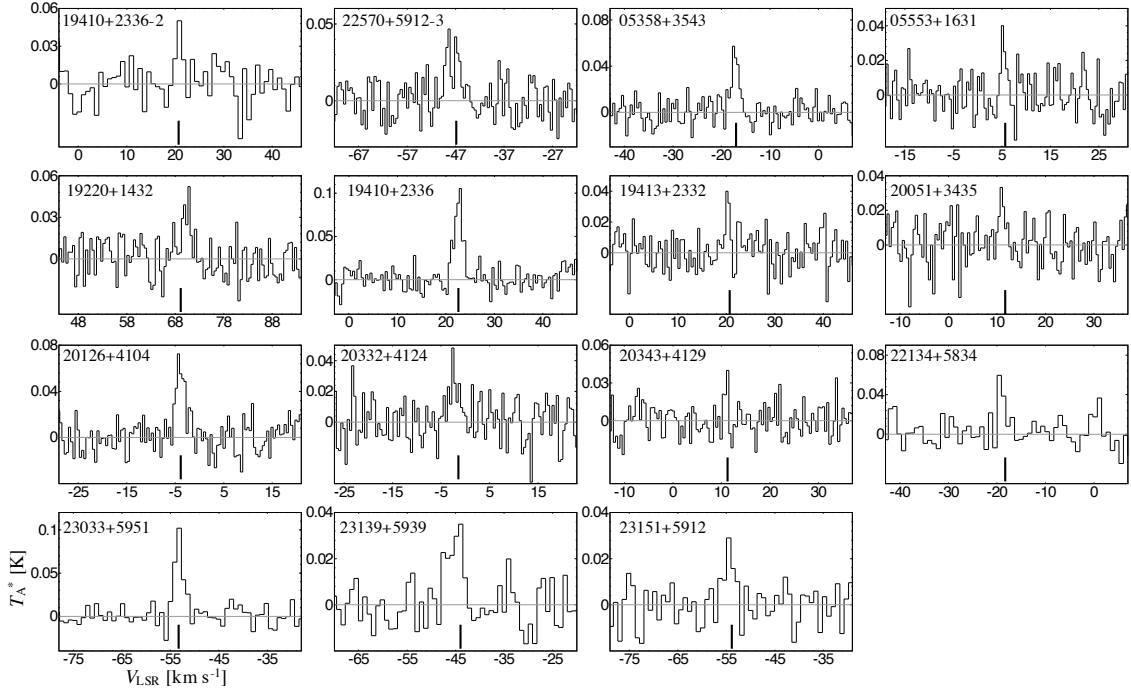}
\caption{Spectra of CCS ($J_{\rm N} =6_{7}-5_{6}$) in HMSCs and HMPOs. The vertical lines indicate the systemic velocities of each source reported by \citep{2002ApJ...566..931S,2005ApJ...634L..57S}. \label{fig:f8}}
\end{figure}

Figures \ref{fig:f7} and \ref{fig:f8} show spectra of {\it c}-C$_{3}$H$_{2}$ and CCS detected with an S/N ratio above 3, including tentative detection, respectively.
The spectral line parameters obtained from the Gaussian fitting are summarized in Table \ref{tab:tab4}.
{\it c}-C$_{3}$H$_{2}$ was detected in 15 HMSCs and 19 HMPOs and CCS was detected in 2 HMSCs and 13 HMPOs.

%%%%%%%%%%%%%%%%%%%%%%%%%%%%%%%%%%%%%%%%%%%%%%%%%%%%%%%%%%%%%%%%%

%%%%%%%%%%%%%%%%%%%%%%%%%%%%%%%%%%%%%%%%%%%%%%%%%%%%%%%%%%%%%%%%%

%%%%%%%%%%%%%%%%%%%%%%%%%%%%%%%%%%%%%%%%%%%%%%%%%%%%%%%%%%%%%%%%%
\floattable
\begin{deluxetable}{lccccccccccc}
\tabletypesize{\scriptsize}
\tablecaption{Spectral line parameters {\it c}-C$_{3}$H$_{2}$ ($J_{\rm Ka, Kc} =2_{0,2} - 1_{1,1}$) and CCS ($J_{\rm N} =6_{7}-5_{6}$) \label{tab:tab4}}
\tablewidth{0pt}
\tablehead{
\colhead{} & \multicolumn{5}{c}{{\it c}-C$_{3}$H$_{2}$ ($J_{\rm Ka, Kc} =2_{0,2} - 1_{1,1}$)} & \multicolumn{5}{c}{CCS ($J_{\rm N} =6_{7}-5_{6}$)} \\
\cline{2-6}\cline{8-12}
\colhead{Source} & \colhead{{\it T}$^{\ast}_{\mathrm A}$} & \colhead{$\Delta v$\tablenotemark{a}} & \colhead{{\it V}$_{\mathrm {LSR}}$\tablenotemark{b}} & \colhead{$\int T^{\ast}_{\mathrm A}dv$} & \colhead{rms\tablenotemark{c}} & \colhead{} & \colhead{{\it T}$^{\ast}_{\mathrm A}$} & \colhead{$\Delta v$\tablenotemark{a}} & \colhead{{\it V}$_{\mathrm {LSR}}$\tablenotemark{b}} & \colhead{$\int T^{\ast}_{\mathrm A}dv$} & \colhead{rms\tablenotemark{c}} \\
\colhead{} & \colhead{(K)} & \colhead{(km s$^{-1}$)} & \colhead{(km s$^{-1}$)} & \colhead{(K km s$^{-1}$)} & \colhead{(mK)} &\colhead{} & \colhead{(K)} & \colhead{(km s$^{-1}$)} & \colhead{(km s$^{-1}$)} & \colhead{(K km s$^{-1}$)} & \colhead{(mK)}
}
\startdata
\multicolumn{11}{l}{HMSCs} \\
	18385-0512-3 & 0.148 (11) & 1.38 (11) & 46.7 & 0.22 (2) & 10.6 & & ... & $< 0.036$ & ... & ... & 11.9 \\
	18437-0216-3 & 0.032 (8) & 1.6 (5) & 111.7 & 0.05 (2) & 10.1 & & $< 0.034$ & ... & ... & ... & 11.4 \\
			      & 0.036 (10) & 0.9 (3) & 97.4 & 0.04 (2) & & & ... & ...& ... & ... & \\
	18445-0222-4 & 0.044 (10) & 2.1 (5) & 88.4 & 0.10 (3) & 10.9 & & $< 0.039$ & ... & ... & ... & 13.1 \\
	18447-0229-4 & 0.055 (12) & 1.7 (4) & 100.5 & 0.10 (3) & 11.1 & & $< 0.039$ & ... & ... & ... & 13.0 \\
	18447-0229-5 & 0.058 (11) & 2.2 (5) & 104.9 & 0.14 (4) & 11.4 & & $< 0.036$ & ... & ... & ... & 11.9 \\
	18454-0158-1 & 0.038 (8) & 3.4 (8) & 99.6 & 0.13 (4) & 10.7 & & $< 0.032$ & ... & ... & ... & 10.7 \\
	18454-0158-3 & 0.047 (7) & 5.9 (9) & 99.5 & 0.30 (6) & 11.8 & & $< 0.034$ & ... & ... & ... & 11.3 \\
	18454-0158-5 & 0.092 (12) & 2.1 (3) & 94.9 & 0.21 (4) & 11.6 & & $< 0.034$ & ... & ... & ... & 11.3 \\
	18454-0158-8 & 0.027 (8) & 3.4 (1.1) & 95.6 & 0.10 (4) & 10.0 & & $< 0.033$ & ... & ... & ... & 11.0 \\
	18454-0158-9 & 0.036 (9) & 3.0 (9) & 101.1 & 0.11 (4) & 10.6 & & $< 0.034$ & ... & ... & ... & 11.3 \\  
			       & 0.038 (11) & 2.0 (7) & 95.7 & 0.08 (4) & 10.6 & & ... & ... & ... & ... & \\ 
	19175+1357-3 & 0.036 (9) & 3.1 (9) & 7.3 & 0.12 (4) & 11.2 & & $< 0.034$ & ... & ... & ... & 11.4 \\
	19175+1357-4 & 0.048 (11) & 1.8 (5) & 7.3 & 0.09 (3) & 12.2 & & $< 0.049$ & ... & ... & ... & 12.3 \\ 
	19410+2336-2 & 0.092 (11) & 2.2 (3) & 21.6 & 0.21 (4) & 12.1 & & 0.051 (14) & 1.5 (4) & 20.8 & 0.08 (3) & 13.7 \\
	20081+2720-1 & 0.045 (7) & 2.0 (4) & 5.7 & 0.10 (2) & 10.9 & & $< 0.032$ & ... & ... & ... & 10.5 \\
	22570+5912-3 & 0.031 (4) & 4.4 (7) & -46.7 & 0.15 (3) & 10.6 & & 0.031 (6) & 3.7 (8) & -48.4 & 0.12 (3) & 11.4 \\ 
	\multicolumn{11}{l}{HMPOs} \\
	05358+3543 & 0.052 (5) & 3.3 (4) & -16.3 & 0.18 (3) & 9.7 & & 0.047 (6) & 2.0 (3) & -17.5 & 0.10 (2) & 9.6 \\
	05490+2658 & 0.046 (9) & 1.4 (3) & 1.0 & 0.07 (2) & 11.5 & & $< 0.037$ & ... & ... & ... & 12.2 \\  
	05553+1631 & $< 0.030$ & ... & ... & ... & 10.0 & & 0.038 (10) & 1.0 (3) & 5.1 & 0.04 (1) & 10.0 \\
	18447-0229 & 0.054 (8) & 1.8 (3) & 102.0 & 0.10 (2) & 11.1 & & $< 0.033$ & ... & ... & ... & 11.1 \\ 
	19035+0641 & 0.026 (4) & 4.9 (9) & 34.1 & 0.14 (3) & 9.8 & & $< 0.035$ & ... & ... & ... & 11.6 \\ 
	19175+1357 & 0.028 (5) & 3.3 (7) & 14.5 & 0.10 (3) & 9.8 & & $< 0.032$ & ... & ... & ... & 10.5 \\ 
	19217+1651 & 0.041 (6) & 4.1 (7) & 1.8 & 0.18 (4) & 11.9 & & $<0.035$ & ... & ... & ... & 11.8 \\ 
	19220+1432 & 0.056 (5) & 4.4 (5) & 70.8 & 0.27 (4) & 10.3 & & 0.035 (7) & 2.4 (6) & 70.8 & 0.09 (3) & 11.6 \\
	19410+2336 & 0.062 (7) & 2.3 (3) & 22.8 & 0.15 (3) & 11.0 & & 0.088 (8) & 2.1 (2) & 23.0 & 0.20 (3) & 11.5 \\  
	19411+2306 & 0.038 (6) & 2.7 (5) & 29.8 & 0.11 (3) & 9.3 & & $< 0.031$ & ... & ... & ... &10.4 \\
	19413+2332 & 0.042 (7) & 2.0 (4) & 20.6 & 0.09 (2) & 10.5 & & 0.040 (10) & 1.0 (3) & 20.1 & 0.04 (2) & 11.5 \\
	20051+3435 & 0.032 (7) & 2.1 (5) & 12.2 & 0.07 (2) & 10.9 & & 0.031 (10) & 1.3 (5) & 10.9 & 0.04 (2) & 11.8 \\
	20126+4104 & 0.219 (7) & 2.45 (10) & -3.3 & 0.57 (3) & 11.8 & & 0.061 (8) & 2.3 (4) & -4.3 & 0.15 (3) & 12.4 \\
	20216+4107 & 0.145 (9) & 1.41 (10) & -1.7 & 0.22 (2) & 10.9 & & $< 0.040$ & ... & ... & ... & 13.3 \\ 
	20293+3952 & 0.063 (7) & 2.6 (3) & 5.6 & 0.17 (3) & 11.3 & & $< 0.040$ & ... & ... & ... & 13.4 \\ 
	20332+4124 & 0.056 (5) & 3.6 (4) & -2.9 & 0.21 (3) & 11.6 & & 0.030 (8) & 2.5 (8) & -2.6 & 0.08 (3) & 13.2 \\
	20343+4129 & 0.059 (8) & 2.6 (3) & 11.3 & 0.16 (3) & 11.0 & & 0.040 (12) & 0.5 (2) & 11.3 & 0.02 (1) & 12.2 \\
	22134+5834 & 0.049 (7) & 3.0 (5) & -17.6 & 0.16 (3) & 10.1 & & 0.059 (13) & 1.2 (3) & -19.5 & 0.08 (3) & 11.7 \\ 
	23033+5951 & 0.059 (8) & 3.1 (5) & -52.8 & 0.19 (4) & 10.5 & & 0.098 (9) & 2.1 (2) & -53.2 & 0.21 (3) & 10.0 \\
	23139+5939 & 0.035 (7) & 2.2 (5) & -43.7 & 0.08 (3) & 8.0 & & 0.031 (8) & 3.2 (9) & -44.0 & 0.10 (4) & 9.6 \\
	23151+5912 & $<0.027$ & ... & ... & ... & 9.3 & & 0.026 (7) & 2.3 (7) & -54.5 & 0.06 (3) & 8.6 \\
\enddata
\tablecomments{The numbers in parentheses represent the standard deviation in the Gaussian fit. The errors are written in units of the last significant digit. The upper limits correspond to $3 \sigma$ values.}
\tablenotetext{a}{These values are not corrected for instrumental velocity resolution.}
\tablenotetext{b}{The error of $V_{\rm {LSR}}$ is commonly 0.8 km s$^{-1}$, corresponding to the velocity resolution of the final spectra (Section \ref{sec:obs}).}
\tablenotetext{c}{The rms noises were evaluated in emission-free region.}
\end{deluxetable}	
%%%%%%%%%%%%%%%%%%%%%%%%%%%%%%%%%%%%%%%%%%%%%%%%%%%%%%%%%%%%%%%%%%%%%%%%%

\section{Analyses} \label{sec:ana}

\subsection{Column Densities and Rotational Temperatures of HC$_{3}$N} \label{sec:anaHC3N}

We derived the rotational temperatures and column densities of HC$_{3}$N from the rotational diagram analysis, using the following formula \citep{1999ApJ...517..209G};
\begin{equation} \label{rd}
{\rm {ln}} \frac{3k \int T_{\mathrm {mb}}dv}{8\pi ^3 \nu S \mu ^2} = {\rm {ln}} \frac{N}{Q(T_{\rm {rot}})} - \frac{E_{\rm {u}}}{kT_{\rm {rot}}},
\end{equation}
where $k$ is the Boltzmann constant, $S$ is the line strength, $\mu$ is the permanent electric dipole moment, $N$ is the column density, and $Q(T_{\rm {rot}})$ is the partition function.
The permanent electric dipole moment is 3.7312 D for HC$_{3}$N \citep{1985JChPh...82...1702D}.
We combined the $J=9-8$ and $10-9$ data with the $J=5-4$ transition (45.49031 GHz, $E_{\rm {u}}/k = 6.5$ K) obtained with the Nobeyama 45-m telescope \citep{2018ApJ...854..133T}.

We derived the column densities assuming two cases.
First, assuming that emission regions of HC$_{3}$N is much larger than a beam size at 45 GHz with the Nobeyama 45-m telescope ($37\arcsec$), we analyzed the data without the beam size correction.
Second, we multiplied the integrated intensities of the $J=5-4$ data by ($\frac{37\arcsec}{18\arcsec}$)$^{2}$ for the beam size correction, assuming small beam filling factors.
We assume that the emission-region size of HC$_{3}$N is $5\arcsec$, which is much smaller than $18\arcsec$. 
We apply this assumption based on previous observations of HC$_{3}$N toward massive young stellar objects \citep{2018ApJ...866..150T}.
We summarize the derived rotational temperatures and column densities by the above two assumptions in Table \ref{tab:tab6}. 
The rotational temperatures in some sources with the beam size correction are lower than a typical value in low-mass starless dark clouds \citep[6.5 K;][]{1992ApJ...392..551S}, which are unlikely results.
These low rotational temperatures possibly imply the non-LTE or opacity effects.
We thus use values without the beam size correction in the following sections.
In some sources, the rotational temperatures derived without the beam size correction are significantly higher (e.g., $T_{\rm {rot}} = 370$ K in HMPO 05553+1631) compared to other sources.
However, the rotational temperatures derived with beam size correction seem to be reasonable (e.g., $T_{\rm {rot}} = 11$ K in HMPO 05553+1631).
This implies that emission regions are small. 

%%%%%%%%%%%%%%%%%%%%%%%%%%%%%%%%%%%%%%%%%%%%%%%%%%%%%%%%%%%%%%%%%
\floattable
\begin{deluxetable}{lccccc}
\tabletypesize{\scriptsize}
\tablecaption{Rotational temperatures and column densities of HC$_{3}$N \label{tab:tab6}}
\tablewidth{0pt}
\tablehead{
\colhead{} & \multicolumn{2}{c}{Without beam-size correction} & \colhead{} & \multicolumn{2}{c}{With beam-size correction} \\
\cline{2-3} \cline{5-6}  
\colhead{Source} & \colhead{$T_{\rm {rot}}$} &\colhead{$N$} & \colhead{} & \colhead{$T_{\rm {rot}}$} &\colhead{$N$} \\
\colhead{} & \colhead{(K)} & \colhead{($\times10^{12}$ cm$^{-2}$)} & \colhead{} & \colhead{(K)} & \colhead{($\times10^{12}$ cm$^{-2}$)}
}
\startdata
\multicolumn{6}{l}{HMSCs} \\
18385-0512-3 & 10 (3) & 4.9 (1.5) & & 5.3 (1.6) & 18 (5 \\
\multicolumn{6}{l}{18437-0216-3} \\
($V_{\rm {LSR}}$ = 111.0 km/s) & 8 (3) & 1.3 (0.4) & & 4.4 (1.3) & 5.7 (1.8)  \\
18445-0222-4 & 13 (4) & 0.47 (0.14) & & 5.4 (1.6) & 1.6 (0.5) \\
18447-0229-4 & 7 (2) & 1.9 (0.6) & & 3.9 (1.2) & 9 (3) \\
18447-0229-5 & 13 (4) & 0.9 (0.3) & & 5.3 (1.6) & 3.4 (1.1) \\
18454-0158-1 & 11 (3) & 1.8 (0.6) & & 5.6 (1.7) & 6 (2) \\
\multicolumn{6}{l}{18454-0158-3} \\
($V_{\rm {LSR}}$ = 97.9 km/s) & 8 (2) & 12 (3) & & 4.6 (1.4) & 50 (15)  \\
18454-0158-5 & 11 (3) & 9 (2) & & 5.5 (1.6) & 33 (10) \\
18454-0158-8 & 10 (3) & 2.7 (0.8) & & 5.4 (1.6) & 10 (3) \\
\multicolumn{6}{l}{18454-0158-9} \\
($V_{\rm {LSR}}$ = 99.5 km/s) & 24 (7) & 9 (3) & & 8 (2) & 22 (7) \\
($V_{\rm {LSR}}$ = 95.9 km/s) & 9 (3) & 10 (3) & & 5.0 (1.5) & 38 (12) \\
19175+1357-3 & 12 (3) & 1.1 (0.3) & & 5.7 (1.7) & 3.9 (1.2) \\
19175+1357-4 &  12 (3) & 1.4 (0.4) & & 5.7 (1.7) & 5.0 (1.5) \\
19410+2336-2 & 22 (7) & 8 (2) & & 7 (2) & 19 (6) \\
22570+5912-3 & 6.3 (1.9) & 0.9 (0.3) & & 3.7 (1.1) & 4.8 (1.5) \\
\multicolumn{6}{l}{HMPOs} \\
05358+3543 & 16 (5) & 9 (3) & & 7 (2) & 28 (8) \\
05490+2658 & 11 (3) & 0.47 (0.15) & & 4.9 (1.5) & 1.9 (0.6) \\
05553+1631 & {\it {370}} & {\it {13.6}} & & 11 (3) & 2.9 (0.9)  \\
18445-0222 & 8 (2) & 2.4 (0.7) & & 4.8 (1.4) & 10 (3) \\
18447-0229 & 12 (4) & 1.0 (0.3) & & 5.9 (1.8) & 3.3 (1.0) \\
19035+0641 & 11 (3) & 4.4 (1.4) & & 5.6 (1.7) & 16 (5) \\
19074+0752 & 25 (7) & 2.4 (0.7) & & 8 (2) & 5.4 (1.7) \\
19175+1357 & {\it {109}} & {\it {4.38}} & & 10 (3) & 2.9 (0.9) \\
19217+1651 & 21 (6) & 11 (3) & & 7 (2) & 27 (8) \\
19220+1432 & 32 (9) & 10 (3) & & 8 (2) & 19 (6)  \\
19266+1745 & 10 (3) & 4.1 (1.3) & & 5.4 (1.6) & 15 (4) \\
19410+2336 & 35 (10) & 23 (7) & & 9 (2) & 39 (12)  \\
19411+2306 & 11 (3) & 2.5 (0.7) & & 5.6 (1.7) & 9 (3) \\
19413+2332 & 9 (3) & 1.5 (0.5) & & 4.9 (1.5) & 6.1 (1.9) \\
20051+3435 & 12 (3) & 1.9 (0.6) & & 5.7 (1.7) & 7 (2) \\
20126+4104 & ... & ... & & 13(4) & 35 (10)  \\
20216+4107 & 51 (15) & 8 (2) & & 9 (3) & 10 (3) \\
20293+3952 & 17 (5) & 6.4 (1.9) & & 7 (2) & 18 (6) \\
20319+3958 & 19 (6) & 1.0 (0.3) & & 7 (2) & 2.7 (0.8) \\
20332+4124 & 12 (4) & 3.2 (0.9) & & 5.9 (1.8) & 11 (3) \\
20343+4129 & 15 (4) & 5.0 (1.5) & & 6.4 (1.9) & 15 (5) \\
22134+5834 & 10 (3) & 8 (2) & & 5.3 (1.6) & 31 (9) \\
22570+5912 & 7 (2) & 1.5 (0.5) & & 4.3 (1.3) & 7 (2) \\
23033+5951 & 24 (7) & 15 (4) & & 8 (2) & 33 (10) \\
23139+5939 & 27 (8) & 7 (2) & & 8 (2) & 15 (5) \\
23151+5912 & 13 (4) & 1.1 (0.3) & & 6.1 (1.8) & 3.5 (1.1) \\
\enddata 
%\tablenotetext{a}{Taken from \citet{2002ApJ...566..931S, 2005ApJ...634L..57S}. The italic letter indicates the average values, which are 15.3 and 22.5 K for HMSCs and HMPOs, respectively.}
\tablecomments{The numbers in parentheses represent the standard deviation.}
\end{deluxetable}	
%%%%%%%%%%%%%%%%%%%%%%%%%%%%%%%%%%%%%%%%%%%%%%%%%%%%%%%%%%%%%%%%%%%%%%%%%

\subsection{Column Densities and Excitation Temperatures of N$_{2}$H$^{+}$} \label{sec:anaN2H+}

We derived the column densities and excitation temperatures of N$_{2}$H$^{+}$ from the LTE analysis.
In case that the spectral line parameters of the three hyperfine components ($F_{1} = 1-1$, $2-1$, and $0-0$) were obtained, we followed the procedure and used formulae derived in Appendix B1 by \citet{2006ApJ...653.1369F}.
For the calculation, we used the values obtained by the Gaussian fitting (Table \ref{tab:t3}).
We assume that all of the hyperfine components have the same thermal and non-thermal line broadenings. 
This can be applicable because all of these lines should come from the same emission region.
We used the line width of $F_{1}=0-0$ as the intrinsic line width, because this component is not blended and consists of only one component.
The derived excitation temperature, optical depth, and column density are summarized in Table \ref{tab:tab5}.
In some sources, we cannot derive its excitation temperatures and column densities from the simultaneous hyperfine-component fitting.
In that case, we derived its column densities using the $F_{1}=2-1$ component (the central strongest line) and the formula (97) in \citet{2015PASP..127..266M}, assuming that its excitation temperature is equal to the average values in HMSCs (4.0 K) and HMPOs (4.7 K), respectively.

The derived excitation temperatures of N$_{2}$H$^{+}$ are lower than the rotational temperatures of NH$_{3}$ \citep{2002ApJ...566..931S, 2005ApJ...634L..57S} or HC$_{3}$N (Table \ref{tab:tab6}).
\citet{2008ApJ...678.1049S} derived the excitation temperatures of N$_{2}$H$^{+}$ in massive clumps.
Their results also show that the excitation temperatures of N$_{2}$H$^{+}$ are lower than the rotational temperatures of NH$_{3}$.
These results seem to suggest that N$_{2}$H$^{+}$ trace colder regions in massive clumps compared to NH$_{3}$.

%%%%%%%%%%%%%%%%%%%%%%%%%%%%%%%%%%%%%%%%%%%%%%%%%%%%%%%%%%%%%%%%%
\floattable
\begin{deluxetable}{lccc}
\tabletypesize{\scriptsize}
\tablecaption{Excitation temperatures, optical depth, and column densities of N$_{2}$H$^{+}$ \label{tab:tab5}}
\tablewidth{0pt}
\tablehead{
\colhead{Source} & \colhead{$T_{\rm {ex}}$} & \colhead{$\tau$} & \colhead{$N$} \\
\colhead{} & \colhead{(K)} & \colhead{} & \colhead{($\times10^{13}$ cm$^{-2}$)}
}
\startdata
 \multicolumn{4}{l}{HMSCs} \\
	18385-0512-3  & 3.9 (0.3) & 6.3 (0.6) & 1.89 (0.16) \\
	18445-0222-4 & 3.1 (0.6) & 8.8 (1.6) & 1.9 (0.3) \\
	18447-0229-3 & {\it 4.0} & & 0.18 (0.04)  \\
	18447-0229-4 & 3.4 (0.5 & 4.0 (0.6) & 0.86 (0.13)  \\	
	18447-0229-5 & 3.2 (0.4) & 8.1 (1.0) & 1.7 (0.2)  \\
	18454-0158-1 & 3.9 (0.3) & 11.0 (0.8) & 5.2 (0.4) \\
	18454-0158-5 & 5.3 (0.3) & 4.4 (0.2) & 3.02 (0.15)  \\
	18454-0158-8 & 4.4 (0.2) & 9.2 (0.5) & 5.9 (0.3) \\	
       18454-0158-10 & 3.3 (0.4) & 4.7 (0.6) & 2.2 (0.3) \\
	19175+1357-3 & 3.3 (0.4) & 6.8 (0.8) & 1.9 (0.2)  \\
	19175+1357-4 & 3.6 (0.7) & 1.9 (0.4) & 0.98 (0.18)  \\
	19410+2336-2 & 7.1 (0.2) & 3.2 (0.1) & 3.62 (0.12) \\
	20081+2720-1 & 3.1 (0.5) & 7.9 (1.2) & 1.3 (0.2)  \\
	22570+5912-3 & 4.7 (0.8) & 0.90 (0.14) & 0.66 (0.11) \\
	\multicolumn{4}{l}{HMPOs} \\
	05358+3543 & 6.8 (0.3) & 3.60 (0.15) & 2.95 (0.12)  \\
	05490+2658 & {\it 4.7} & & 0.14 (0.07)  \\
	05553+1631 & 3.7 (0.8) & 0.80 (0.17) & 0.28 (0.06)  \\
	18445-0222 & 3.1 (0.5) & 2.6 (0.4) & 1.15 (0.18) \\
	18447-0229 & 3.8 (0.3) & 6.7 (0.5) & 2.24 (0.18) \\
	19035+0641 & 3.7 (0.3) & 3.5 (0.2) & 2.66 (0.18)\\
	19074+0752 & 5.2 (0.6) & 0.70 (0.08) & 0.51 (0.05) \\
	19175+1357 & {\it 4.7} & & 0.60 (0.05) \\
	19217+1651 & 5.3 (0.2) & 2.1 (0.1) & 2.99 (0.11) \\
	19266+1745 & 3.6 (0.3) & 2.4 (0.2) & 1.41 (0.11) \\
	19282+1814 & 3.0 (1.1) & 4.9 (1.7) & 0.9 (0.3) \\
	19410+2336 & 10.1 (0.3) & 5.10 (0.16) & 7.6 (0.2) \\
	19411+2306 & 4.3 (0.4) & 3.3 (0.3) & 1.14 (0.11) \\
	19413+2332 & 3.1 (0.4) & 6.2 (0.8) & 1.7 (0.2) \\
	20051+3435 & 3.1 (0.4) & 7.6 (1.1) & 2.1 (0.3) \\
	20126+4104 & 9.6 (0.2) & 3.60 (0.07) & 6.04 (0.12) \\
	20205+3948 &  {\it 4.7} & & 0.05 (0.02) \\
	20216+4107 & 5.0 (0.2) & 6.2 (0.3) & 2.37 (0.10) \\
	20293+3952 & 5.2 (0.4) & 2.3 (0.2) & 2.16 (0.18) \\
	20319+3958 & 3.1 (0.7) & 4.1 (0.9) & 0.84 (0.19) \\
	20332+4124 & 3.6 (0.2) & 4.6 (0.3) & 2.00 (0.14) \\
	20343+4129 & 5.3 (0.2) & 4.30 (0.17) & 2.27 (0.09) \\
	22134+5834 & 3.7 (0.5) & 2.4 (0.3) & 0.85 (0.11) \\
	22570+5912 & 3.1 (1.2) & 2.3 (0.9) & 0.40 (0.16) \\
	23033+5951 & 6.7 (0.3) & 2.40 (0.11) & 3.07 (0.14) \\
	23139+5939 & 4.3 (0.3) & 3.7 (0.3) & 1.66 (0.12) \\
\enddata
\tablecomments{The numbers in parentheses represent the standard deviation. We cannot derive $T_{\rm {ex}}$ and $N$ from the LTE analysis in some sources. We derived $N$ with fixed $T_{\rm {ex}}$ at the average values in HMSCs (4.0 K) and HMPOs (4.7 K), respectively, shown as the italic letter.}
\end{deluxetable}	
%%%%%%%%%%%%%%%%%%%%%%%%%%%%%%%%%%%%%%%%%%%%%%%%%%%%%%%%%%%%%%%%%%%%%%%%%

\section{Discussion} \label{sec:dis}

\subsection{Comparisons of Line Widths} \label{sec:disline}

\begin{figure}
\figurenum{9}
\plotone{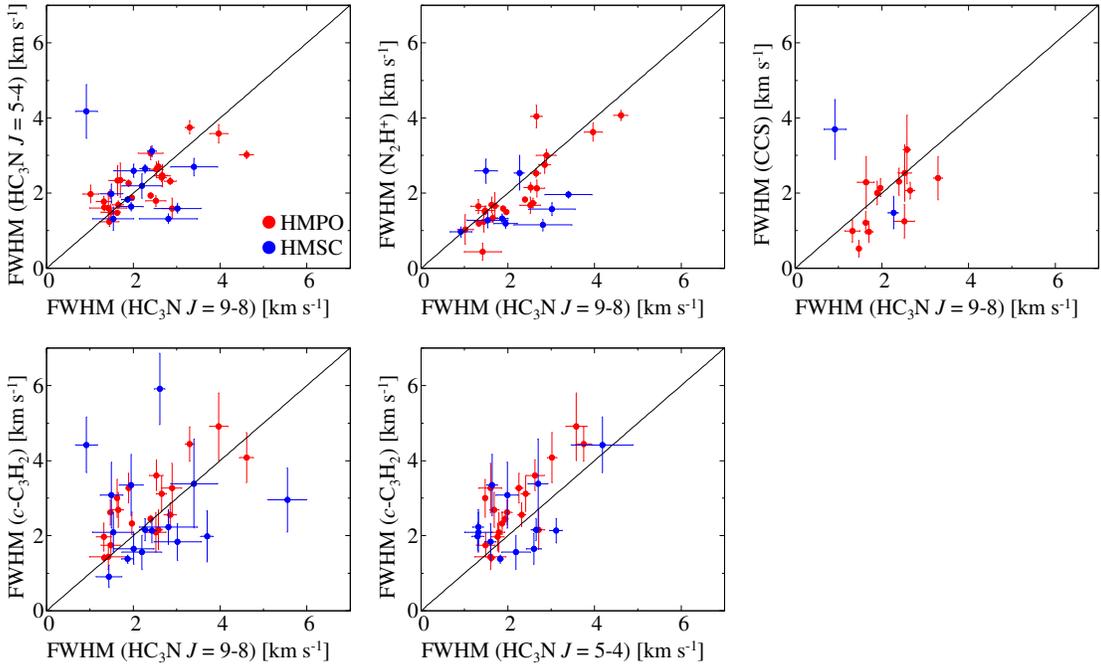}
\caption{Relationship of FWHM among observed molecular lines. The black lines indicate the FWHM (species a) = FWHM (species b). The error bars show the standard deviation. \label{fig:f13}}
\end{figure}

We compare line widths of the observed molecular lines, including the $J=5-4$ transition of HC$_{3}$N \citep[45.49031 GHz, $E_{\rm {u}}/k = 6.5$ K;][]{2018ApJ...854..133T}, with the $J=9-8$ transition of HC$_{3}$N as shown in Figure \ref{fig:f13}.
We compare the $J=5-4$ transition of HC$_{3}$N and that of $c$-C$_{3}$H$_{2}$ because their excitation energies are similar to each other.

The line widths of HC$_{3}$N are similar and have a good correlation between the two transitions.
This means that these two transitions trace similar regions; dense gas regions \citep{2018ApJ...854..133T}.
The line widths of N$_{2}$H$^{+}$ and CCS are similar to and correlated with that of HC$_{3}$N, suggesting that they trace similar regions with HC$_{3}$N.
Our result of a correlation of the line widths between HC$_{3}$N and N$_{2}$H$^{+}$ is consistent with a work by \citet{2008ApJ...678.1049S}.

$c$-C$_{3}$H$_{2}$, on the other hand, shows slightly larger line widths compared to those of HC$_{3}$N, as shown the plot of FWHM($c$-C$_{3}$H$_{2}$) vs. FWHM(HC$_{3}$N $J=5-4$).
In the panel of FWHM($c$-C$_{3}$H$_{2}$) vs. FWHM(HC$_{3}$N $J=9-8$), the plots are scattered.
These may mean that $c$-C$_{3}$H$_{2}$ trace inner warmer regions, compared to other carbon-chain species.
\citet{2001PASJ...53..251T} also showed that $c$-C$_{3}$H$_{2}$ emission is strong at the central dense core of the protostar in L1527.

We also compared the mean and median values between HMSCs and HMPOs as listed in Table \ref{tab:FWHM}.
We do not recognize any differences in line width between HMSC and HMPO.

%%%%%%%%%%%%%%%%%%%%%%%%%%%%%%%%%%%%%%%%%%%%%%%%%%%%%%%%%%%%%%%%%
\floattable
\begin{deluxetable}{lccccc}
\tabletypesize{\scriptsize}
\tablecaption{Comparisons of the mean and median values of FWHM between HMSCs and HMPOs \label{tab:FWHM}}
\tablewidth{0pt}
\tablehead{
\colhead{} & \colhead{HC$_{3}$N ($J =9-8$)} & \colhead{N$_{2}$H$^{+}$} & \colhead{CCS} & \colhead{{\it {c}}-C$_{3}$H$_{2}$} & \colhead{HC$_{3}$N ($J=5-4$)} 
}
\startdata
\multicolumn{6}{l}{Average}	 \\						
HMSC &	2.50 & 1.82 & 2.59 & 2.53 & 2.33 \\
HMPO &	2.23 & 1.94 & 1.83 & 2.80 & 2.17 \\		
\multicolumn{6}{l}{Median}	\\
HMSC & 2.27 & 1.96 & 2.59 & 2.14 	& 2.09 \\
HMPO & 2.18 & 1.66 & 2.07 & 2.62 	& 2.05 \\		
\enddata
\tablecomments{The unit for this table values is km s$^{-1}$.}
\end{deluxetable}	
%%%%%%%%%%%%%%%%%%%%%%%%%%%%%%%%%%%%%%%%%%%%%%%%%%%%%%%%%%%%%%%%%%%%

%%%%%%%%%%%%%%%%%%%%%%%%%%%%%%%%%%%%%%%%%%%%%%%%%%%%%%%%%%%%%%%%%
%%%%%%%%%%%%%%%%%%%%%%%%%%%%%%%%%%%%%%%%%%%%%%%%%%%%%%%%%%%%%%%%%

\subsection{A Possible Chemical Evolutionary Indicator, $N$(N$_{2}$H$^{+}$)/$N$(HC$_{3}$N)} \label{sec:evolution}

In this paper, we test the $N$(N$_{2}$H$^{+}$)/$N$(HC$_{3}$N) column density ratio for a chemical evolutionary indicator in high-mass star-forming regions, motivated by \citet{1992ApJ...392..551S} and \citet{2009ApJ...699..585H}.
The different source distances and/or different source sizes could cause uncertainties in the column density estimate.
The column density ratios, however, are robust, if we assume that the spatial distributions of each species are similar in all the sources.
Since HC$_{3}$N is known as an early-type carbon-chain species in low-mass star-forming regions \citep[e.g.,][]{1992ApJ...392..551S} and detected from almost all of the sources, we choose HC$_{3}$N as an early-type species.
We select N$_{2}$H$^{+}$ as a late-type species because it has confirmed in low-mass star-forming regions \citep{1998ApJ...506..743B}. 
In addition, we can avoid significant beam dilution and contamination from other sources as far as possible; N$_{2}$H$^{+}$ can be observed with a smaller beam size compared to NH$_{3}$, e.g., the ($J, K$) = (1,1) inversion transition in the 23 GHz band is observed with the beam size of $\sim 80 \arcsec$ with the Nobeyama 45-m telescope.
Both HC$_{3}$N and N$_{2}$H$^{+}$ seem to trace similar regions (dense core parts) as discussed in Section \ref{sec:disline}.
Therefore, we will study the chemical evolution of massive cores, using the $N$(N$_{2}$H$^{+}$)/$N$(HC$_{3}$N) ratio.
Although we use different molecules from \citet{1992ApJ...392..551S} and \citet{2009ApJ...699..585H}, N$_{2}$H$^{+}$ and HC$_{3}$N have been known as a late-type species and an early-type species in low-mass star-forming regions, as mentioned in Section \ref{sec:intro}.
Therefore, we expect that the $N$(N$_{2}$H$^{+}$)/$N$(HC$_{3}$N) ratio increases from HMSCs to HMPOs, if the ratio in high-mass star-forming regions similarly changes.

\begin{figure}
\figurenum{10}
\plotone{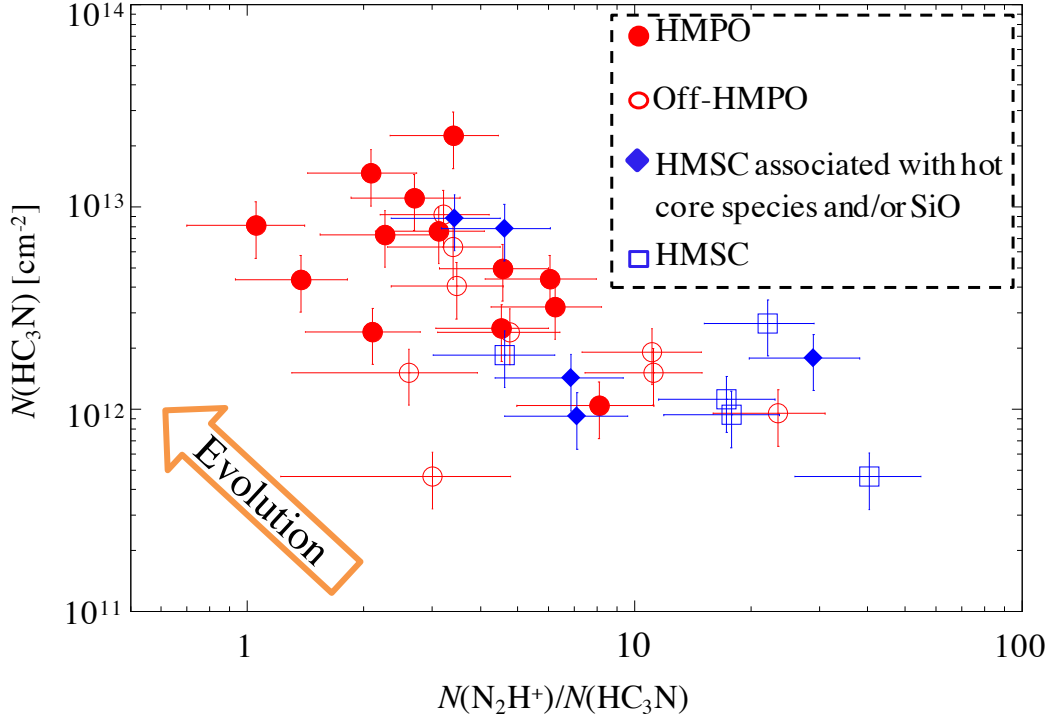}
\caption{Relationship between $N$(N$_{2}$H$^{+}$)/$N$(HC$_{3}$N) and $N$(HC$_{3}$N). The error bars show the standard deviation. Off-HMPO means that IRAS observed positions were not at exact continuum peak positions, but the beam covered the continuum core in the beam edge. \label{fig:f9}}
\end{figure}

Figure \ref{fig:f9} shows the relationship between the $N$(N$_{2}$H$^{+}$)/$N$(HC$_{3}$N) ratio and $N$(HC$_{3}$N) in high-mass star-forming regions.
We checked the 1.2 mm continuum images of target sources \citep{2002ApJ...566..945B, 2005ApJ...634L..57S} and found that several IRAS observed positions were not at exact continuum peak positions, but the beam covered the continuum core in the beam edge.
We plot such data points as ``Off-HMPO" in Figure \ref{fig:f9}.
In some HMSCs, saturated COMs (CH$_{3}$OH and CH$_{3}$CN) and/or SiO were detected \citep{2007ApJ...668..348B}. 
These molecular emission lines indicate star formation activity \citep{2007ApJ...668..348B}.
We plot these HMSCs as ``HMSC associated with hot core species and/or SiO".

Figure \ref{fig:f9} shows that the $N$(N$_{2}$H$^{+}$)/$N$(HC$_{3}$N) ratio tends to decrease from HMSCs to HMPOs and the $N$(HC$_{3}$N) value increases. 
As we mentioned in Section \ref{sec:intro}, the $N$(nitrogen-bearing molecules)/$N$(carbon-chain molecules) ratio increases from starless cores to star-forming cores in low-mass star-forming regions.
Hence, this tendency in high-mass star-forming regions is clearly opposite to that in low-mass star-forming regions \citep{1992ApJ...392..551S, 2009ApJ...699..585H}.

We conducted the Kolmogorov-Smirnov (K-S) test about the $N$(N$_{2}$H$^{+}$)/$N$(HC$_{3}$N) ratio and $N$(HC$_{3}$N).
We included samples of ``HMPO" and ``HMSC".
We summarize the probabilities that the $N$(N$_{2}$H$^{+}$)/$N$(HC$_{3}$N) ratios and the $N$(HC$_{3}$N) values in HMSCs and HMPOs originate from the same parent populations in Table \ref{tab:KS}.
The probabilities were derived to be 0.75\% and 1.54\% for the $N$(N$_{2}$H$^{+}$)/$N$(HC$_{3}$N) ratio and the $N$(HC$_{3}$N) value, respectively.

Furthermore, we conducted the Welch's t test.
The results are summarized in Table \ref{tab:KS}.
The average value of the $N$(N$_{2}$H$^{+}$)/$N$(HC$_{3}$N) ratio in HMSCs (20.4) is higher than that in HMPOs (3.7).
The average value of $N$(HC$_{3}$N) in HMSCs is lower than that in HMPOs by a factor of 5.
Thus, the differences in the $N$(N$_{2}$H$^{+}$)/$N$(HC$_{3}$N) ratio and the $N$(HC$_{3}$N) value between HMSC and HMPO are reliable.

%%%%%%%%%%%%%%%%%%%%%%%%%%%%%%%%%%%%%%%%%%%%%%%%%%%%%%%%%%%%%%%%%
%%%%%%%%%%%%%%%%%%%%%%%%%%%%%%%%%%%%%%%%%%%%%%%%%%%%%%%%%%%%%%%%%

%%%%%%%%%%%%%%%%%%%%%%%%%%%%%%%%%%%%%%%%%%%%%%%%%%%%%%%%%%%%%%%%%
\floattable
\begin{deluxetable}{lcc}
\tabletypesize{\scriptsize}
\tablecaption{Summary of statistical analyses \label{tab:KS}}
\tablewidth{0pt}
\tablehead{
\colhead{} & \colhead{$N$(N$_{2}$H$^{+}$)/$N$(HC$_{3}$N)} & \colhead{$N$(HC$_{3}$N)}
}
\startdata
K-S test p-value & 0.75\% & 1.54\% \\
Welch's t test p-value & 4.36\% & 0.43\% \\
Mean (HMSC)  & 20.4 & $1.4 \times 10^{12}$ cm$^{-2}$ \\
Mean (HMPO) & 3.7 & $7.2 \times 10^{12}$ cm$^{-2}$ \\
\enddata
\end{deluxetable}	
%%%%%%%%%%%%%%%%%%%%%%%%%%%%%%%%%%%%%%%%%%%%%%%%%%%%%%%%%%%%%%%%%%%%

\subsection{Explanation for the Changes in the $N$(N$_{2}$H$^{+}$)/$N$(HC$_{3}$N) Ratio} \label{sec:d2}

\begin{figure}
\figurenum{11}
\plotone{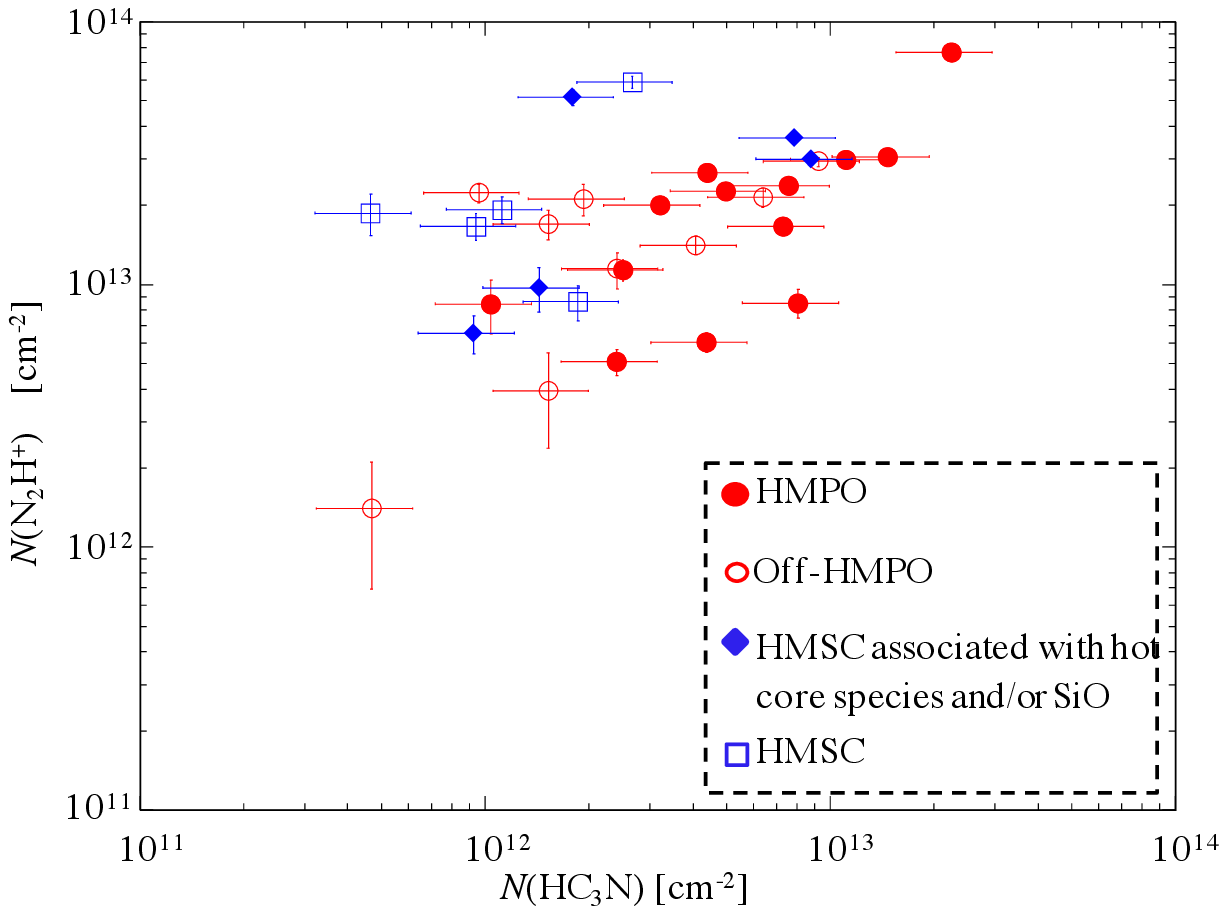}
\caption{Relationship between $N$(N$_{2}$H$^{+}$) and $N$(HC$_{3}$N). The error bars show the standard deviation. \label{fig:f10}}
\end{figure}

Two possible factors can contribute to the decrease in the $N$(N$_{2}$H$^{+}$)/$N$(HC$_{3}$N) ratio: an increase in $N$(HC$_{3}$N) and a decrease in $N$(N$_{2}$H$^{+}$).
Figure \ref{fig:f10} shows the plot of $N$(HC$_{3}$N) vs. $N$(N$_{2}$H$^{+}$).

%%%%%%%%%%%%%%%%%%%%%%%%%%%%%%%%%%%%%%%%%%%%%%%%%%%%%%%%%%%%%%%%%
%%%%%%%%%%%%%%%%%%%%%%%%%%%%%%%%%%%%%%%%%%%%%%%%%%%%%%%%%%%%%%%%%

Plots of N$_{2}$H$^{+}$ in HMPOs seem to slightly below those in HMSCs.
We conducted the K-S test and Welch's t test for the N$_{2}$H$^{+}$ column density, $N$(N$_{2}$H$^{+}$).
The probability that the $N$(N$_{2}$H$^{+}$) values in HMSCs and HMPOs originate from the same parent populations was derived to be 77\%.
The average values of $N$(N$_{2}$H$^{+}$) are $2.4 \times 10^{13}$ cm$^{-2}$ and $2.3 \times 10^{13}$ cm$^{-2}$ in HMSCs and HMPOs, respectively.
Hence, the N$_{2}$H$^{+}$ column density does not significantly change from HMSCs to HMPOs.

Taking the fact that the excitation temperatures of N$_{2}$H$^{+}$ (Table \ref{tab:tab5}) are lower than the rotational temperatures of HC$_{3}$N (Table \ref{tab:tab6}) or NH$_{3}$ \citep[$T_{\rm {rot}}\sim 20$ K;][]{2002ApJ...566..931S} into consideration, N$_{2}$H$^{+}$ seems to survive in colder parts of cores.
Besides the electron recombination reactions, N$_{2}$H$^{+}$ is destroyed by the reaction with CO molecules, which are sublimated from grain mantles with the dust temperature above 20 K \citep{1983A&A...122..171Y}, as follows:
\begin{equation} \label{reaction:CO}
{\rm {N}}_{2}{\rm {H}}^{+} + {\rm {CO}} \rightarrow {\rm {N}}_{2} + {\rm {HCO}}^{+}
\end{equation}
The dust temperatures in HMPOs are much higher than 20 K \citep[$T_{\rm {cd}}\sim 50$ K;][]{2002ApJ...566..931S}.
Hence, N$_{2}$H$^{+}$ may have been already destroyed in HMPOs at some degree, but it is not clear in our sample.
This may be caused by the single-dish observation covering large linear scales.
However, the reaction (\ref{reaction:CO}) is still one of the possible explanations for the low excitation temperatures of N$_{2}$H$^{+}$.

As discussed in Section \ref{sec:evolution} and summarized in Table \ref{tab:KS}, the HC$_{3}$N column density increases from HMSCs to HMPOs.
HC$_{3}$N may be formed from CH$_{4}$ \citep{2008ApJ...681.1385H} and/or C$_{2}$H$_{2}$ \citep{2009MNRAS.394..221C} evaporated from grain mantles in HMPOs \citep{2016ApJ...830..106T,2018ApJ...854..133T}.
The sublimation temperatures of CH$_{4}$ and C$_{2}$H$_{2}$ are approximately 25 K and 50 K, respectively \citep{1983A&A...122..171Y}.
The dust temperatures in HMPOs are higher than these sublimation temperatures, and thus they are possible parent species of HC$_{3}$N.

\begin{figure}
\figurenum{12}
\plotone{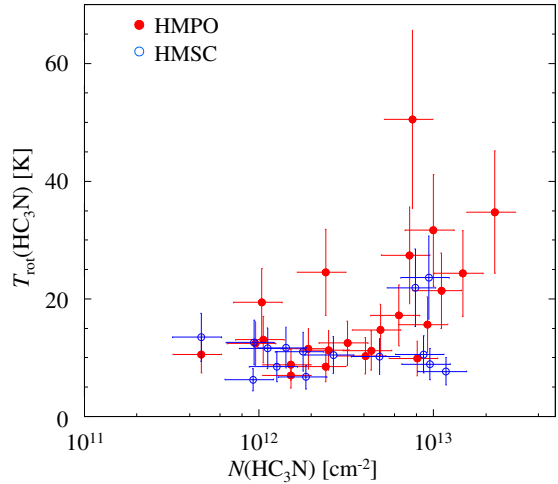}
\caption{Relationship between rotational temperature and column density of HC$_{3}$N. The error bars show the standard deviation. \label{fig:f11}}
\end{figure}

We investigated the relationship between the column density and rotational temperature of HC$_{3}$N, as shown in Figure \ref{fig:f11}.
We conducted the Kendall's rank correlation statics.
The probability that $N$(HC$_{3}$N) and $T_{\rm {rot}}$(HC$_{3}$N) are not related is 62.6\% in HMSCs.
The corresponding probability and the Kendall's tau correlation coefficient ($\tau$) in HMPOs are 0.65\% and $+0.41$, respectively.
Hence, $N$(HC$_{3}$N) tends to increase with  $T_{\rm {rot}}$(HC$_{3}$N) in HMPOs, whereas they are not related with each other in HMSCs.
This may suggest that HC$_{3}$N is newly formed around massive young protostars.

As discussed above, the gas-phase chemical composition around massive young protostars may be affected by the molecules evaporated from grain mantles; CO, CH$_{4}$, and/or C$_{2}$H$_{2}$.
This could bring the difference between high-mass star-forming regions and low-mass star-forming regions.
The dust temperatures around massive young protostars are higher than those around low-mass protostars and the wider area should be heated by massive young protostars.
In such a condition in high-mass star-forming regions, more molecules are sublimated from grain mantles in wider regions and the sublimated species may have more significant effects on the gas-phase chemical reactions.
 
%%%%%%%%%%%%%%%%%%%%%%%%%%%%%%%%%%%%%%%%%%%%%%%%%%%%%%%%%%%%%%%%%
%%%%%%%%%%%%%%%%%%%%%%%%%%%%%%%%%%%%%%%%%%%%%%%%%%%%%%%%%%%%%%%%%

\subsection{Comparisons of Detection Rates of Carbon-Chain Species with Previous Studies} \label{sec:detection_rate}

As we mentioned before (Section \ref{sec:res}), this dataset is the deepest survey observations of carbon-chain molecules in high-mass star-forming regions.
We then discuss including weak molecular emission lines in this subsection.

The detection rates of HC$_{3}$N, $c$-C$_{3}$H$_{2}$, and CCS are 93, 68, and 46\%, respectively, in HMPOs, including tentative detection.
\citet{2018ApJ...854..133T} reported that the detection rate of HC$_{5}$N is 50\% in HMPOs.

\citet{2018ApJ...863...88L} carried out observations toward 16 deeply embedded (Class 0/I) low-mass protostars using the IRAM 30-m telescope.
They reported that the detection rates of CCS, CCCS, HC$_{3}$N, HC$_{5}$N, $l$-C$_{3}$H, and C$_{4}$H are 88\%, 38\%, 75\%, 31\%, 81\%, and 88\%, respectively.

We found that the detection rates of carbon-chain molecules are probably different between high-mass protostars and low-mass protostars.
In our sample, HC$_{3}$N has been detected in almost all of our target HMPOs, but HC$_{5}$N and CCS have been detected in approximately half of sample.
On the other hand, CCS is most frequently detected and HC$_{5}$N shows the lowest detection rate around low-mass protostars.
The low detection rate of HC$_{5}$N around low-mass protostars may be caused by observed transitions; \citet{2018ApJ...863...88L} observed HC$_{5}$N using the high-excitation-energy lines ($E_{\rm {u}}/k > 80.5$ K), while \citet{2018ApJ...854..133T} observed its $J=16-15$ line ($E_{\rm {u}}/k = 17.4$ K).
However, the results that HC$_{3}$N is more commonly detected around high-mass protostars are plausible.

Some differences between chemistry of cyanopolyynes and CCS may cause the different detection rates between high-mass star-forming regions and low-mass star-forming regions.
For example, cyanopolyynes and $c$-C$_{3}$H$_{2}$ do not react with atomic oxygen in the gas phase\footnote{We investigated using UMIST 2012 (http://udfa.ajmarkwick.net/index.php) and Kinetic Database for Astrochemistry (KIDA; http://kida.obs.u-bordeaux1.fr)}, whereas CCS reacts with oxygen atoms to produce CO and CS.
Some possibilities for the abundant atomic oxygen in high-mass star-forming regions could be considered:
\begin{itemize}
\item Atomic oxygen is efficiently evaporated from grain mantles with the dust temperature above $\sim 60$ K \citep[e.g.,][]{2013ApJ...779...11F}.
\item Strong UV radiation destroys CO molecules and the abundance of atomic oxygen may be high around HMPOs, compared to around low-mass protostars \citep{2006A&A...449..609J,2016Natur.537..207G}.
\item Cosmic rays may contribute to the formation of atomic oxygen via the following reactions \citep[e.g.,][]{2017A&A...605A..57F}:
\begin{equation} \label{reaction:destCO}
{\rm {He}}^{+} + {\rm {CO}} \rightarrow {\rm {O}} + {\rm {C}}^{+} + {\rm {He}}.
\end{equation}
\end{itemize}
The second and third ones heavily depend on physical conditions, especially the density structures, and we cannot conclude that they have large impacts.
On the other hand, the first one is the most important origin when the dust temperature is above 60 K, according to the gas-grain-bulk three-phase chemical network simulation (Taniguchi et al., in prep.).
The dust temperatures in HMPOs are higher than 60 K \citep{2002ApJ...566..931S}, and then atomic oxygen can be evaporated from dust grains.
CCS may be efficiently destroyed around HMPOs by the reaction with atomic oxygen and the detection rate decreases.

Further observations and chemical network simulations are necessary for explanations for differences in carbon-chain chemistry between high-mass protostars and low-mass protostars.

\section{Conclusions}

We have carried out survey observations of molecular lines of HC$_{3}$N, N$_{2}$H$^{+}$, $c$-C$_{3}$H$_{2}$, and CCS in the 81$-$94 GHz band toward 17 HMSCs and 28 HMPOs with the Nobeyama 45-m radio telescope.
We achieved the deepest survey observations of carbon-chain species in high-mass star-forming regions.

We investigate the $N$(N$_{2}$H$^{+}$)/$N$(HC$_{3}$N) ratio as a chemical evolutionary indicator in high-mass star-forming regions.
The $N$(N$_{2}$H$^{+}$)/$N$(HC$_{3}$N) ratio decreases from HMSCs to HMPOs, which is an opposite result to low-mass star-forming regions.
From the statistical analyses, we confirmed that the HC$_{3}$N column density increases from HMSCs to HMPOs, whereas that of N$_{2}$H$^{+}$ does not significantly change.
In addition, the excitation temperatures of N$_{2}$H$^{+}$ are lower than the rotational temperatures of HC$_{3}$N and NH$_{3}$ in both HMSCs and HMPOs.
This implies that N$_{2}$H$^{+}$ exists in cold dense cores, where the CO molecules do not sublimate from grain mantles.
On the other hand, HC$_{3}$N may be newly formed from CH$_{4}$ and/or C$_{2}$H$_{2}$ evaporated from grain mantles in the warm gas around massive young protostars.
This seems to be supported by the correlation between the rotational temperature and column density of HC$_{3}$N in HMPOs.

We compare the detection rates of carbon-chain species between high-mass protostars and low-mass protostars.
In high-mass protostars, the detection rates of cyanopolyynes are higher compared to low-mass protostars, while the detection rate of CCS is significantly low in high-mass protostars.
We discuss one possible interpretation involving atomic oxygen for the different carbon-chain detection rates between high-mass protostars and low-mass protostars.
Additional observations and chemical network simulations are necessary.

%% Putting eqnarrays or equations inside the mathletters environment groups
%% the enclosed equations by letter. For instance, the eqnarray below, instead
%% of being numbered, say, (4) and (5), would be numbered (4a) and (4b).
%% LaTeX the paper and look at the output to see the results.

\acknowledgments

We are deeply grateful to the staff of the Nobeyama Radio Observatory.
The Nobeyama Radio Observatory is a branch of the National Astronomical Observatory of Japan (NAOJ), National Institutes of Natural Science (NINS).
We would like to express our heartfelt appreciation to Professor Paola Caselli (Max Planck Institute for Extraterrestrial Physics) for her convictive advices.
K. T. appreciates support from a Granting-Aid for Science Research of Japan (17J03516).
K. T. would like to thank the University of Virginia for providing the funds for her postdoctoral fellowship in the VICO research program.

\vspace{5mm}
\facilities{Nobeyama 45-m radio telescope}


\software{Java Newstar}



\appendix

\section{Rotational Diagram Fitting of HC$_{3}$N}

%%%%%%%%%%%%%%%%%%%%%%%%%%%%%%%%%%%%%%%%%%%%%%%%%%%%%%%%%%%%%%%%%

\begin{figure}
\figurenum{13}
\plotone{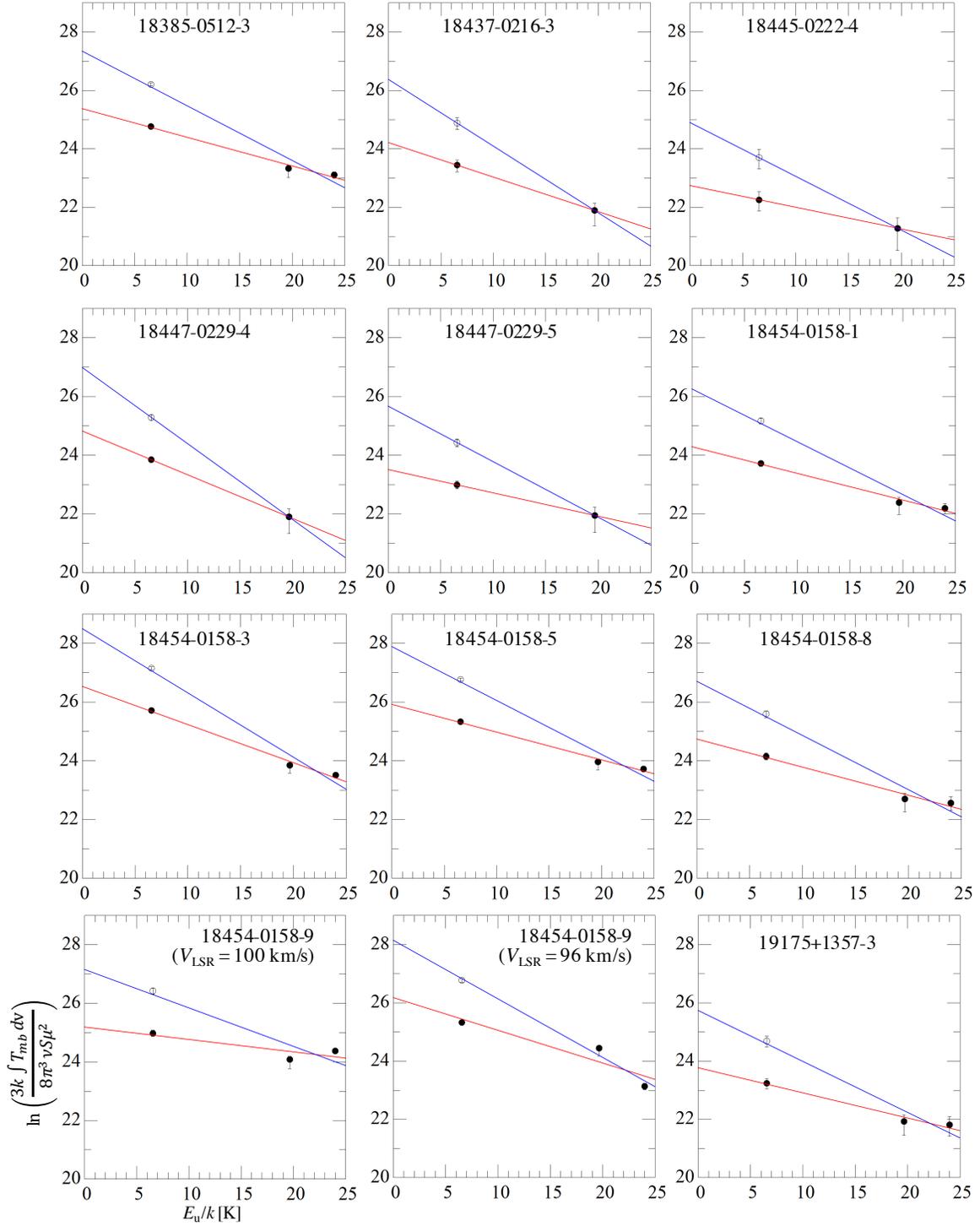}
\caption{Rotational diagram of HC$_{3}$N. Open circle and filled circle of $J=5-4$ transition represent values with and without beam size correction, respectively. Blue and red lines indicate fitting results for with and without beam size correction, respectively. The errors were derived from the Gaussian fitting of spectra. \label{fig:a1}}
\end{figure}

\begin{figure}
\figurenum{14}
\plotone{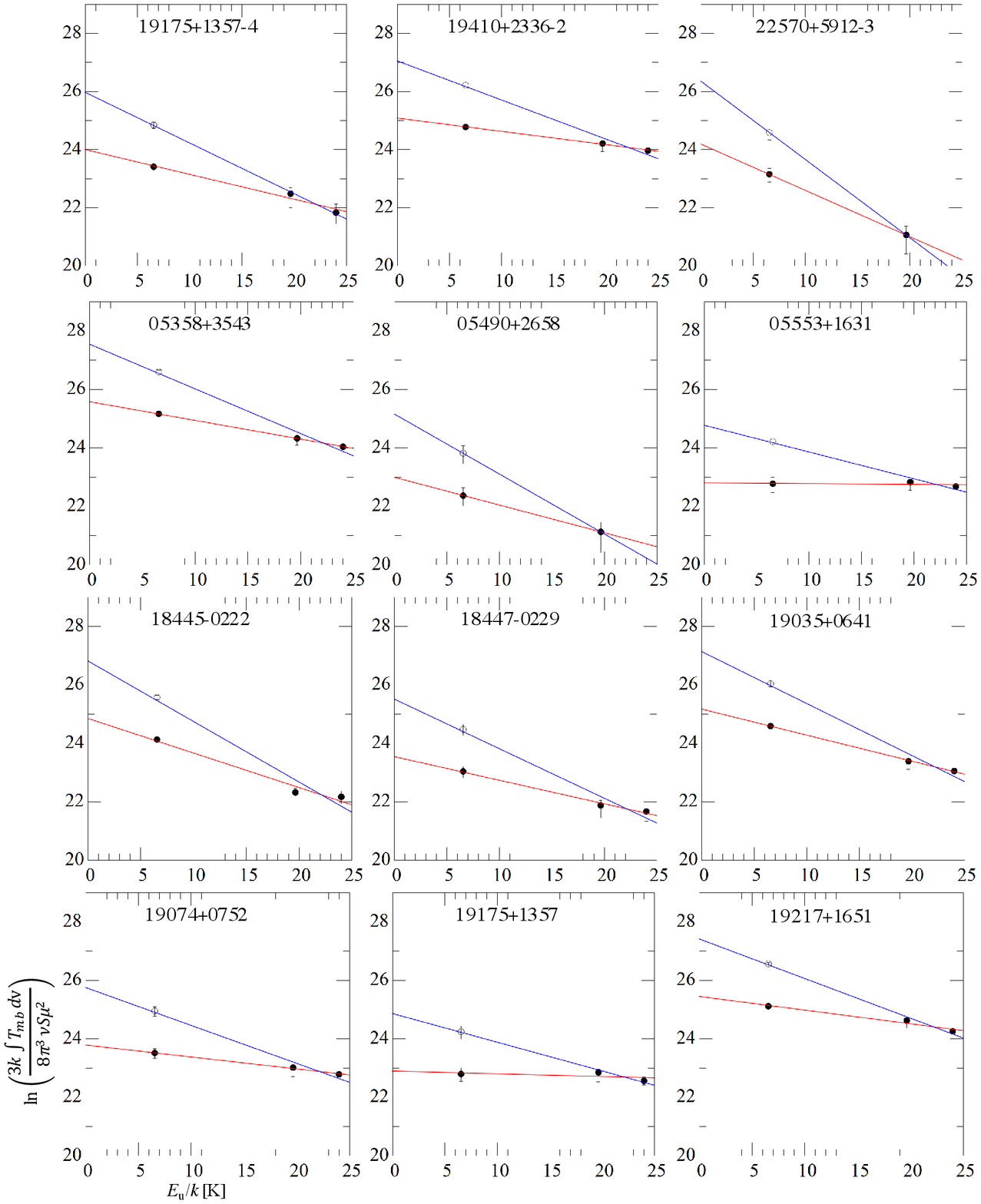}
\caption{Continue. \label{fig:a2}}
\end{figure}

\begin{figure}
\figurenum{15}
\plotone{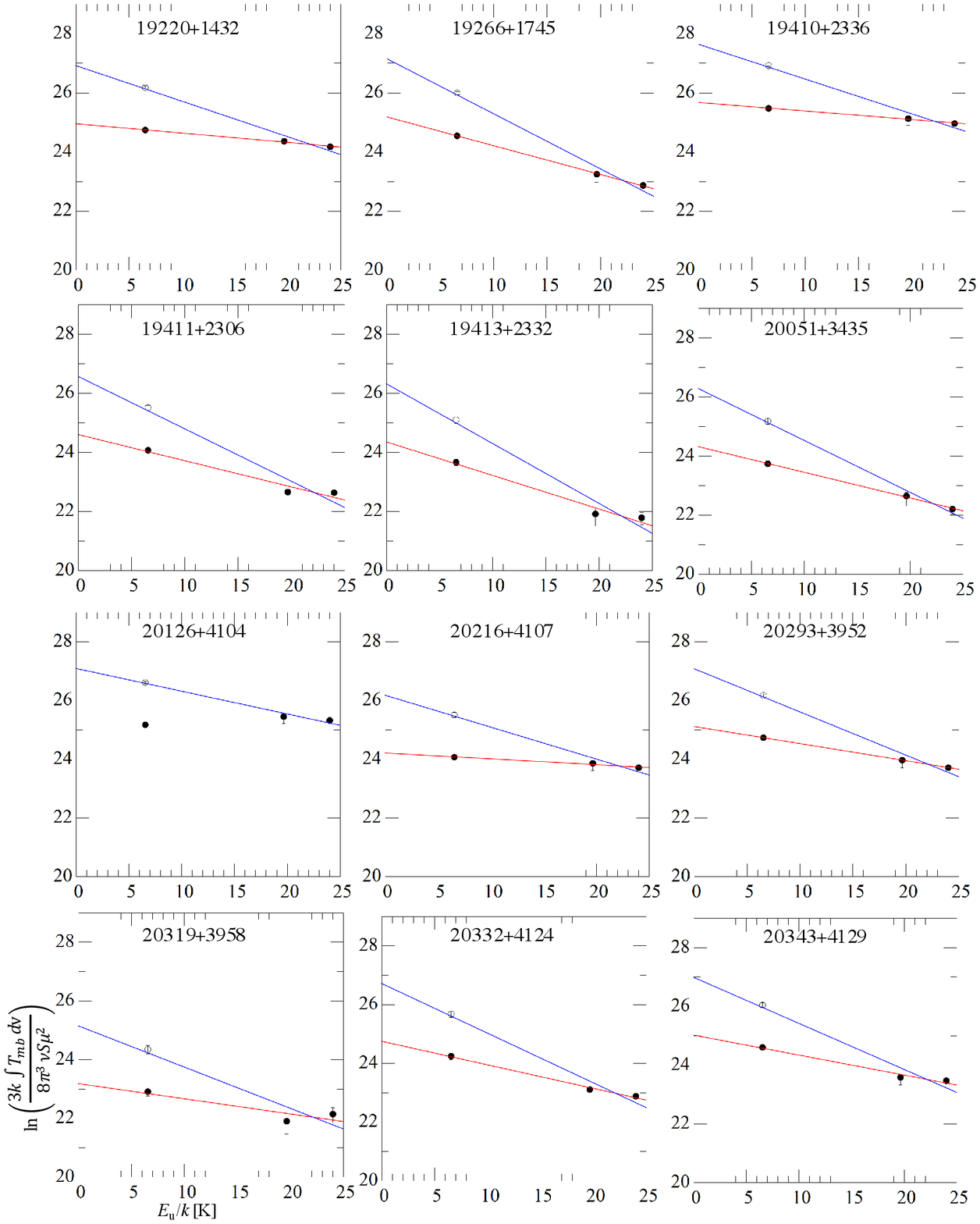}
\caption{Continue. \label{fig:a3}}
\end{figure}

\begin{figure}
\figurenum{16}
\plotone{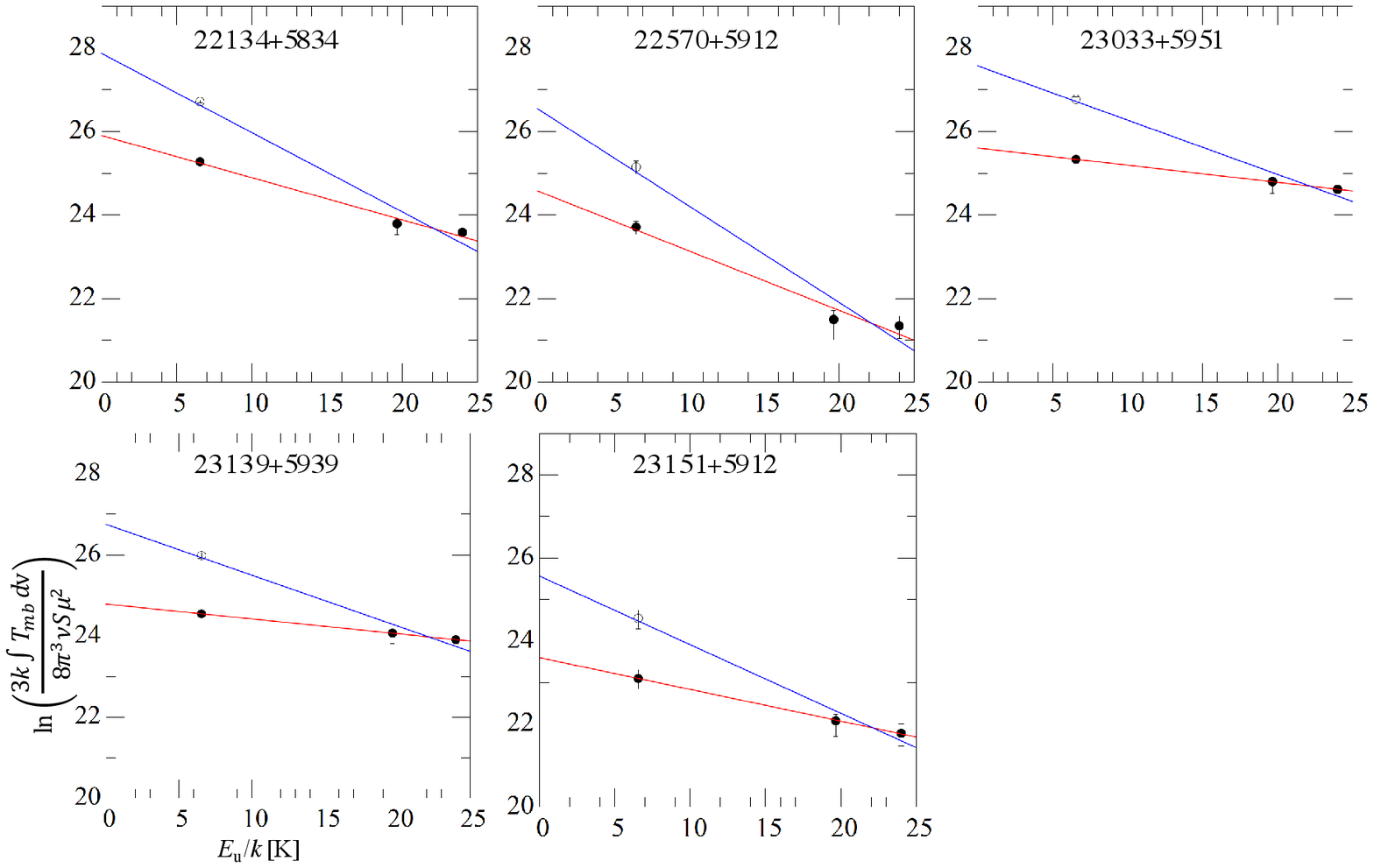}
\caption{Continue. \label{fig:a4}}
\end{figure}

\end{document}